\documentclass[aip,amsmath,amssymb,floatfix,reprint]{revtex4-1}

\usepackage[english]{babel}
\usepackage{graphicx}
\usepackage{float}
\usepackage{dcolumn}
\usepackage{bm}
\usepackage{braket}
\usepackage[usenames,dvipsnames]{color}
\usepackage{ragged2e}
\usepackage[labelsep=period, format=plain, justification=justified, font=footnotesize]{caption}
\usepackage{times}
\usepackage{txfonts}
\usepackage{listings}
\usepackage{xcolor}
\usepackage{placeins}
\usepackage[version=4]{mhchem}

\definecolor{codegreen}{rgb}{0,0.6,0}
\definecolor{codegray}{rgb}{0.5,0.5,0.5}
\definecolor{codepurple}{rgb}{0.58,0,0.82}
\definecolor{backcolour}{rgb}{0.95,0.95,0.92}

\lstdefinestyle{mystyle}{
    backgroundcolor=\color{backcolour},
    commentstyle=\color{codegreen},
    keywordstyle=\color{magenta},
    numberstyle=\tiny\color{codegray},
    stringstyle=\color{codepurple},
    basicstyle=\ttfamily\footnotesize,
    breakatwhitespace=false,
    breaklines=true,
    captionpos=b,
    keepspaces=true,
    numbers=left,
    numbersep=5pt,
    showspaces=false,
    showstringspaces=false,
    showtabs=false,
    tabsize=2
}
\usepackage{microtype}
\usepackage{hyperref}

\hypersetup{colorlinks = true,
pdfauthor = {Artrith},
linkcolor = blue,
urlcolor  = blue,
citecolor = blue,
anchorcolor = blue]{hyperref}
}

\makeatletter
\let\ORIbbl@fixname\bbl@fixname
\def\bbl@fixname#1{%
  \@ifundefined{languagealias@\expandafter\string#1}
    {\ORIbbl@fixname#1}
    {\edef\languagename{\@nameuse{languagealias@#1}}}%
}
\newcommand{\definelanguagealias}[2]{%
  \@namedef{languagealias@#1}{#2}%
}
\makeatother
\definelanguagealias{en}{english}

\DeclareCaptionJustification{myjust}{\justifying}
\begin{document}

\title{AENET-LAMMPS and AENET-TINKER: Interfaces for Accurate and Efficient Molecular Dynamics Simulations with Machine Learning Potentials}

\author{Michael S. Chen}
\thanks{These authors contributed equally}
\affiliation{
Department of Chemistry, Stanford University, Stanford, CA 94305, United States
}
\author{Tobias Morawietz}
\thanks{These authors contributed equally}
\affiliation{
Department of Chemistry, Stanford University, Stanford, CA 94305, United States
}
\author{Hideki Mori}
 \affiliation{Department of Mechanical Engineering, College of Industrial Technology, 1-27-1 Nishikoya, Amagasaki, Hyogo 661-0047, Japan
}
\author{Thomas E. Markland}
\email{tmarkland@stanford.edu}
\affiliation{
Department of Chemistry, Stanford University, Stanford, CA 94305, United States
}
\author{Nongnuch Artrith}
\email{n.artrith@uu.nl}
\affiliation{
Department of Chemical Engineering, Columbia University, New York, NY 10027, United States
}
\affiliation{
Columbia Center for Computational Electrochemistry, Columbia University, New York, NY 10027, United States
}
\affiliation{
 Materials Chemistry and Catalysis, Debye Institute for Nanomaterials Science, Utrecht University, 3584 CG Utrecht, The Netherlands
}

\date{\today}

\begin{abstract}
Machine learning potentials (MLPs) trained on data from quantum-mechanics based first-principles methods can approach the accuracy of the reference method at a fraction of the computational cost.
To facilitate efficient MLP-based molecular dynamics (MD) and Monte Carlo (MC) simulations, an integration of the MLPs with sampling software is needed.
Here we develop two interfaces that link the \textbf{a}tomic \textbf{e}nergy \textbf{net}work (\ae{}net) MLP package with the popular sampling packages TINKER and LAMMPS.
The three packages, \ae{}net, TINKER, and LAMMPS, are free and open-source software that enable, in combination, accurate simulations of large and complex systems with low computational cost that scales linearly with the number of atoms.
Scaling tests show that the parallel efficiency of the \ae{}net-TINKER interface is nearly optimal but is limited to shared-memory systems.
The \ae{}net-LAMMPS interface achieves excellent parallel efficiency on highly parallel distributed memory systems and benefits from the highly optimized neighbor list implemented in LAMMPS.
We demonstrate the utility of the two MLP interfaces for two relevant example applications, the investigation of diffusion phenomena in liquid water and the equilibration of nanostructured amorphous battery materials.
\end{abstract}
\maketitle
\normalsize


\section{Introduction}

Atomistic simulations based on molecular dynamics (MD) or Monte Carlo (MC) techniques~\cite{frenkel2001,allen2017} have become standard tools for the \textit{in silico} characterization and prediction of materials and molecular properties.
Due to their ability to include realistic experimental conditions, such as solvent effects, temperature, and pressure, they are routinely employed in a wide range of research areas in both academic and industrial settings.
Prominent examples are the identification of novel materials for energy applications~\cite{curtarolo2013,jain2016,urban2016,seh2017,oganov2019} or the design of novel drugs~\cite{jorgensen2004,vandrie2007,aminpour2019,morawietz_Machine_2020}.
To perform these simulations, reliable interatomic potentials (or \textit{force fields}) are required that describe the atomic interactions~\cite{becker2013,jorgensen2005}.
They need to be accurate to capture the atomistic process of interest but also efficient to generate simulations on sufficient time and length scales using reasonable compute resources.
Machine-learning potentials (MLPs)~\cite{behler2007,bartok2010,artrith2011,jose_construction_2012,botu2015,artrith2016,shapeev_Moment_2016,khorshidi2016,s.smith2017,unke2018,schutt2018,mori_Neural_2020,miksch_Strategies_2021}, including approaches based on artificial neural networks (ANNs), learn the interatomic interactions from accurate quantum mechanical (QM) calculations such as density-functional theory (DFT)~\cite{burke2012} and have shown great promise in combining accuracy and affordability allowing to simulate complex systems under realistic conditions\cite{artrith2011, artrith_neural_2013, artrith2019, mueller2020, morawietz_Machine_2020}.

In this work, we discuss interfaces of two popular freely available simulation codes, the TINKER molecular modeling software~\cite{ponder1987} and the \emph{Large-scale Atomic/Molecular Massively Parallel Simulator} (LAMMPS)~\cite{plimpton1995}, with the \textbf{a}tomic \textbf{e}nergy \textbf{net}work (\ae{}net) ANN potential package~\cite{artrith2016}.
In combination, these tools enable routine long-time simulations of large and complex systems with computational cost that scales linearly with system size and can be distributed across compute cores without significant loss of efficiency.

We note that there are already several MLP methods available in LAMMPS, for example, the Spectral Neighbor Analysis Potential (SNAP)~\cite{thompson2015} and the Gaussian Approximation Potential (GAP)~\cite{bartok2010} approaches, each with their own strengths and weaknesses. Potentials trained with the GAP approach provide an intrinsic estimate of the prediction error.
An advantage of the SNAP method is its computational efficiency for simple (especially elemental) compounds.
The Behler-Parrinello ANN potential method~\cite{behler2007} that is subject of the present work is computationally more demanding but can often achieve greater accuracy~\cite{zuo2020}.
Another advantage of the ANN potential method is that it can be trained on large reference data sets with millions of samples, complex structures, and many chemical elements~\cite{s.smith2017}.

In the following section (Sec.~\ref{sec:methods}) we briefly review the ANN potential method.
The basic functionality of the simulation framework and the required input for incorporating \ae{}net ANN interatomic potentials are described in Sec.~\ref{sec:implementation}.
Section~\ref{sec:motivation} illustrates the need for computationally efficient simulation tools with accurate ANN potentials as demonstrated for two examples of complex atomic systems: amorphous lithium-silicon (a-LiSi) alloys and liquid water.
The computational efficiency of both interfaces is discussed in Sec.~\ref{sec:performance}.
All code interfaces presented here are freely available and shared, together with the corresponding input files, user instructions, and ANNs for LiSi~\cite{artrith_constructing_2018,artrith2019a} and liquid water~\cite{morawietz2018,morawietz2019}.

\section{Methods}
\label{sec:methods}

\subsection{The artificial neural network (ANN) potential method}

\noindent
A feedforward ANN is a vector function that can be expressed as the recurrence relation
\begin{align}
  \mathbf{x}^{l}
  = f_{\textup{a}}^{l}\bigl(
    \mathbf{w}^{l} \cdot \mathbf{x}^{l-1} + \mathbf{b}^{l}
  \bigr)
  \quad ,
  \label{eq:ANN-layer}
\end{align}
where the values of the neurons $\mathbf{x}^{l}$ in the $(l)$th layer depend on the values of the neurons $\mathbf{x}^{l-1}$ in the $(l-1)$th layer, and the elements of matrix $\mathbf{w}^{l}$ and vector $\mathbf{b}^{l}$ are parameter sets called \emph{weights} and \emph{biases}, respectively.
$f_{\textup{a}}^{l}$ is the \emph{activation function} of layer $l$, which acts on each component of its input vector.
It has been shown that ANNs of the form of equation~\eqref{eq:ANN-layer} can represent any function with arbitrary accuracy if suitable non-linear activation functions are used~\cite{cybenko_approximation_1989}.

The initial layer of the ANN, i.e., $\mathbf{x}^{0}$ in Eq.~\eqref{eq:ANN-layer}, is given by the function input (\emph{input layer}).
The final \emph{output layer} of the ANN corresponds to the function value.
The layers between the input and output do not have any intuitive meaning and are often referred to as \emph{hidden layers}.
Together, the dimensions of all layers define the \emph{architecture} of the ANN, e.g., $n_{\textup{in}}$-$n_{\textup{h},1}$-$n_{\textup{h},2}$-$n_{\textup{out}}$ is the architecture of an ANN with input dimension $n_{\textup{in}}$, output dimension $n_{\textup{out}}$ and two hidden layers with dimensions $n_{\textup{h},1}$ and $n_{\textup{h},2}$.

Behler and Parrinello~\cite{behler2007} introduced a technique for the representation of interatomic potentials with ANNs by expressing the total energy $E$ of an atomic structure as the sum of atomic energies $E_{i}$
\begin{align}
  E
  = \sum_{i}^{\textup{atoms}} E_{i}
  \approx \sum_{t}^{\textup{types}} \sum_{i}^{\textup{atoms}}
    \textup{ANN}_{t}\Bigl(
      \widetilde{\sigma}_{i}^{R_{\textup{c}}}
    \Bigr)
  \quad ,
\end{align}
where $\textup{ANN}_{t}$ is an ANN trained to predict the atomic energies of atoms with type/species $t$.
The ANN input $\widetilde{\sigma}_{i}^{R_{\textup{c}}}$ is a descriptor vector of the local atomic environment of atom $i$ within a cutoff radius of $R_{\textup{c}}$.

\subsection{Descriptors of the local atomic environment}
\label{sec:descriptor}

\noindent
The potential energy is invariant with respect to rotation and translation of the entire structure as well as to the permutation of equivalent atoms, and the descriptor $\widetilde{\sigma}_{i}^{R_{\textup{c}}}$ also needs to satisfy these symmetries.
Additionally, the size of the descriptor vector determines the dimension of the ANN input layer and must not vary with the number of atoms in an atomic structure.
Therefore, the Cartesian atomic coordinates cannot be directly used as ANN inputs and need first be transformed into an invariant representation.

In the present work, we use the descriptor by Artrith, Urban, and Ceder~\cite{artrith2017}, in which the radial and angular distribution functions (RDF and ADF) of the local atomic environment are expanded in a basis set of Chebyshev polynomials.
The expansion coefficients are a suitable invariant descriptor of the local atomic environment and can be used as the inputs for the ANN potential.
The coefficients of the RDF expansion are given by
\begin{align}
  c^{\textup{pair}}_{\alpha}
  = \sum_{j \ne i} T_{\alpha}\left(
    \frac{2r_{ij}}{R_c} -1
  \right) f_c(r_{ij})
  \quad ,
  \label{eq:radial}
\end{align}
where $r_{ij}$ is the atomic distance between atoms $i$ and $j$, $R_c$ is the cutoff radius, and $T_{\alpha}$ is the Chebyshev polynomial of the first kind with order $\alpha$~\cite{artrith2017}.
The Chebyshev polynomials $T_n$ are defined by a recurrence relation
\begin{align}
  T_{n+1}(x) = 2xT_{n}(x) - T_{n-1}(x)
  \quad ,
\end{align}
where $T_0(x)$ and $T_1(x)$ are 1 and $x$, respectively.
We use a cosine cutoff function defined as
\begin{align}
  f_c(r_{ij}) = \left\{
  \begin{array}{ll}
    \frac{1}{2}\left[
      \cos \left( \frac{\pi r_{ij}}{R_c} \right) + 1
    \right]~~~
    & (r_{ij} \le R_c) \\
    0~~~ & (r_{ij} > R_c)
  \end{array}
  \right.
  \quad .
  \label{eq:cos-cutoff}
\end{align}
The coefficients from the expansion of the ADF are similarly given by
\begin{align}
  c^{\textup{triple}}_{\alpha}
  = \sum_{j \ne i, k \ne i,j} T_{\alpha} \left(
    \cos \theta_{ijk}
  \right) f_c(r_{ij}) f_c(r_{ik})
  \quad ,
  \label{eq:angular}
\end{align}
where $r_{ij}$, $r_{ik}$, and $r_{jk}$ are the atomic distances between atoms $i$, $j$, and $k$, and $\theta_{ijk}$ is the angle defined by the three atoms.

\subsection{ANN potential training}

\noindent
Training an ANN potential requires the optimization of the weight and bias parameters $\mathbf{w}^{l}$ and $\mathbf{b}^{l}$ of equation~\eqref{eq:ANN-layer} to reproduce reference energies as closely as possible.
In the present work, this was achieved by minimizing the cost function $C$ defined as follows
\begin{align}
  C\left(\left\{\mathbf{w},\mathbf{b}\right\}\right)
  = \frac{1}{2}\sum^{N_{\rm s}}_{t=1}\left[
    E^{\rm ANN}_{t}\left(\left\{\mathbf{w},\mathbf{b}\right\}\right)
    - E^{\rm DFT}_{t}
  \right]^2
  \quad ,
\end{align}
where $N_{\rm s}$ is the number of structures in the \emph{training set}, and $E^{\rm ANN}_t$ and $E^{\rm DFT}_t$ are potential energies of the $t$th structure of the ANN potential and the reference DFT calculation, respectively.
The limited memory Broyden-Fletcher-Goldfarb-Shanno (LM-BFGS) method was used for the actual optimization~\cite{broyden_bfgs_1970, fletcher_bfgs_1970, goldfarb_bfgs_1970, shanno_bfgs_1970, liu_lbfgs_1989}, and the gradients of the cost function, $\frac{\partial C}{\partial \mathbf{w}}$ and $\frac{\partial C}{\partial \mathbf{b}}$, were obtained using the conventional back-propagation technique~\cite{lecun_efficient_2012}.

\section{Implementation}
\label{sec:implementation}

\begin{figure*}[tbp]
\centering
  \includegraphics[scale=0.5]{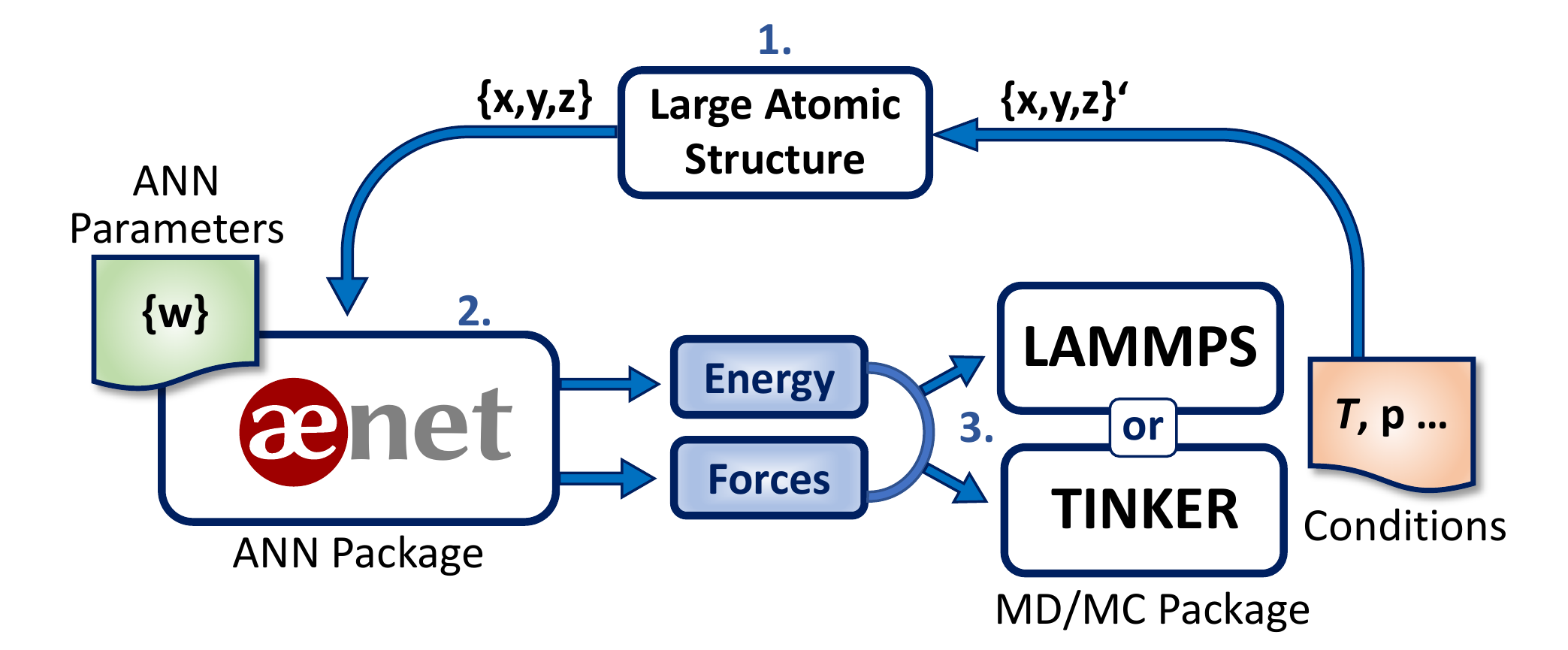}\\
\caption{
\label{fig:interface_overview}
\textbf{Schematic overview of the \ae{}net-LAMMPS / \ae{}net-TINKER program interfaces for efficient atomistic simulations with artificial neural network (ANN) potentials.} The Cartesian coordinates $\bigl\{x,y,z\bigr\}$ of a large (periodic or non-periodic) atomic system (1) are mapped to total energy and atomic forces by a previously trained ANN potential (2) defined by a set of model parameters $\bigl\{w\bigr\}$. Via an interface (3) between the open-source ANN code \ae{}net~\cite{artrith2016}, using the \ae{}net library \textbf{(\ae{}netLib)}, and the molecular dynamics (MD) / Monte Carlo (MC) codes LAMMPS~\cite{plimpton1995} or TINKER~\cite{ponder1987}, force and/or energy information are used to propagate the initial atomic positions to a new structure with coordinates $\bigl\{x,y,z\bigr\}'$ under a set of specified thermodynamic conditions such as temperature, $T$, and pressure, p. The cycle repeats until a trajectory of sufficient length has been generated from which the properties of interest can be obtained.
}
\end{figure*}

\noindent
The \ae{}net package~\cite{artrith2016} is a free and open source software for the construction (training) and evaluation of ANN-based machine-learning potentials.
\ae{}net implements a variant of the high-dimensional neural network potential method by Behler and Parrinello~\cite{behler2007} described in the previous section~\ref{sec:methods}.
The cost of computing \ae{}net’s structure descriptors, detailed in section~\ref{sec:descriptor}, does not scale with the number of atomic species.
This enables the systematic construction of efficient and accurate ANN potentials without manual parametrization for compositions with a large number of atom types~\cite{artrith2017}.
With the Chebyshev descriptor the cumbersome setup of hand-crafted descriptors is avoided and replaced by just a single line of code in the \ae{}net input.

The \ae{}net package can be compiled into a library (\ae{}netLib) that is compatible with the Python, C/C\texttt{++}, and FORTRAN programming languages, facilitating the integration with simulation software.
A Python interface to the Atomic Simulation Environment (ASE)~\cite{larsen_Atomic_2017} is provided in Ref.~\citenum{artrith2016}.
ASE is a Python framework for atomistic simulations and provides a simple API together with calculators for interfacing with third-party software for the evaluation of structural energies and atomic forces.
The \ae{}net package includes an implementation of an ASE calculator linked to \ae{}netLib.
A C\texttt{++} interface to the LAMMPS code is also provided in Ref~\citenum{mori_Neural_2020}.
This interface is designed for the metallurgical analysis in structural materials, such as BCC iron.
For example, this interface extends \ae{}netLib so that the stress tensor of the entire system can be calculated using the functionality in LAMMPS~\cite{thompson2009general}.

In this work, we describe two interfaces with \textit{aenetLib} coupled to software code written in Fortran and C++: the TINKER and LAMMPS codes, respectively.
These simulation software packages are designed with different use cases in mind:
TINKER is implemented in the FORTRAN programming language and is structured into small command-line utilities, each with distinct purpose (e.g., MD simulations, geometry optimizations, MC simulations, etc.).
From a user perspective, TINKER is comparatively easy to learn and install from source code, which makes it a popular choice for desktop or laptop computers.
The standard distribution of TINKER is \emph{shared memory} parallelized with the Open Multi-Processing (OpenMP)~\cite{dagum_OpenMP_1998} standard, so that simulations can make optimal use of modern consumer-grade processors.
LAMMPS, on the other hand, is a highly optimized MD software designed for massively parallel high-performance computer (HPC) systems.
LAMMPS input files are scriptable, which adds flexibility but also steepens the learning curve.
LAMMPS is \emph{distributed-memory} parallelized using the message passing interface (MPI) standard and can scale to thousands of distributed processor cores.
We note that a TINKER distribution for HPC systems has recently been released, named Tinker-HP~\cite{lagardere_TinkerHP_2018, adjoua_TinkerHP_2021}.
Data reported here was, however, obtained with the standard TINKER distribution.

Fig.~\ref{fig:interface_overview} shows the general workflow for performing efficient atomistic simulations with MLPs: The Cartesian coordinates of an initial atomic structure are mapped to their corresponding total energy and atomic forces via the MLP package \ae{}net given an MLP that is parametrized for all elements in the system.
Employing an interface coupling the \ae{}netLib library to either the TINKER or LAMMPS simulation package, the system of interest is propagated to a new set of coordinates under a specified set of thermodynamic conditions.
Both LAMMPS and standard TINKER were interfaced with \ae{}net by linking against the \textit{\ae{}netLib} library, so that \ae{}net routines are used for evaluating structural energies and interatomic forces.
In the following section, we describe the implementation and usage of these interfaces.

Note that \ae{}net-TINKER uses the neighbor list from \ae{}net, which is implemented by a comparatively simple algorithm, while \ae{}net-LAMMPS makes direct use of the highly optimized LAMMPS neighbor list.

\subsection{Building TINKER with \ae{}net support}

Three new files were created to connect the \ae{}net libraries with TINKER: \textit{aenettinker.f90},
\textit{extra.f}, and \textit{extra1.f}.

Using an ANN potential with the TINKER MD code requires to copy the following files to the TINKER source directory,
\begin{lstlisting}
aenettinker.f90
extra.f
extra1.f
\end{lstlisting}
and to compile the TINKER source code using the provided \emph{Makefile}, e.g.,
\begin{lstlisting}
Makefile.aenetlib_ifort
Makefile.aenetlib_gfortran.
\end{lstlisting}

\subsection{Running TINKER simulations using \ae{}net potentials}

To run simulations using \ae{}net ANN potentials, the TINKER parameter file \emph{aenet.prm} is used. It only contains the masses of all chemical species, as no further information is required.

All ANN potential files have to be present in the working directory and need to follow the naming convention

\begin{lstlisting}[language=bash]
<species>.ann,
\end{lstlisting}

where $<$species$>$ is the chemical symbol (H, He, Li, etc.).
In the principal TINKER key input file, \ae{}net potentials are activated with the keyword

\begin{lstlisting}[language=bash]
EXTRATERM only.
\end{lstlisting}

The number of threads used in parallel runs can be controlled with the keyword \emph{OPENMP-THREADS}.
We refer also to the example Tinker input file \emph{tinker.key} on our shared repository.

\subsection{Building LAMMPS with \ae{}net support}

Similar to the TINKER package, two new files,
\begin{lstlisting}
pair_aenet.h
pair_aenet.cpp,
\end{lstlisting}
were written as part of the USER-AENET package to build \ae{}net-LAMMPS by connecting with the \ae{}net libraries.

For building the \ae{}net-LAMMPS interface, the USER-AENET package has to be copied into the LAMMPS source directory and the default Makefile in the LAMMPS source directory needs to be replaced by the one we provide.
The user needs to include this package when building LAMMPS by calling,
\begin{lstlisting}
make yes-user-aenet
\end{lstlisting}
in the LAMMPS source directory and then compile LAMMPS using the Makefile.

\subsection {Running LAMMPS simulations using \ae{}net potentials}

The \ae{}net library files, as well as any other dependencies, need to be properly loaded (i.e. by setting the \$LD$\_$LIBRARY$\_$PATH). The LAMMPS input script also needs to be configured so as to use the \ae{}net \emph{pair style} and to specify which neural network parameter (\emph{.ann}) files to use.

A partial input example for water is provided below:

\begin{lstlisting}[language=bash]
units metal
mass 1  1.007825
mass 2 15.999491
pair_style aenet H.ann O.ann
pair_coeff * *
\end{lstlisting}

The user must specify the order of the \emph{$<$species$>$.ann} potential files such that they correspond to the mass values defined above.
In the example above, element 1 is designated to be hydrogen and element 2 is oxygen.
Consequently, the \ae{}net parameter file for hydrogen (\emph{H.ann}) needs to be specified first and then followed by the parameter file for oxygen (\emph{O.ann}).

The selected LAMMPS unit system should match the units of the reference data used for training the MLP by \ae{}net.
In the example above \emph{metal} units are used, meaning that the \ae{}net potentials were fitted to training data for which the energies were reported in electron-volts and the positions in Angstroms.
If instead the training data is provided in units of Hartrees and Bohrs, then the corresponding LAMMPS unit system \emph{electron} should be chosen.
Further details on the unit systems can be looked up in the LAMMPS manual.

\section{Results and Discussion}
\label{sec:discussion}
\subsection{Complex systems require efficient and accurate simulation tools}
\label{sec:motivation}

\begin{figure*}
\centering
  \includegraphics[scale=0.55]{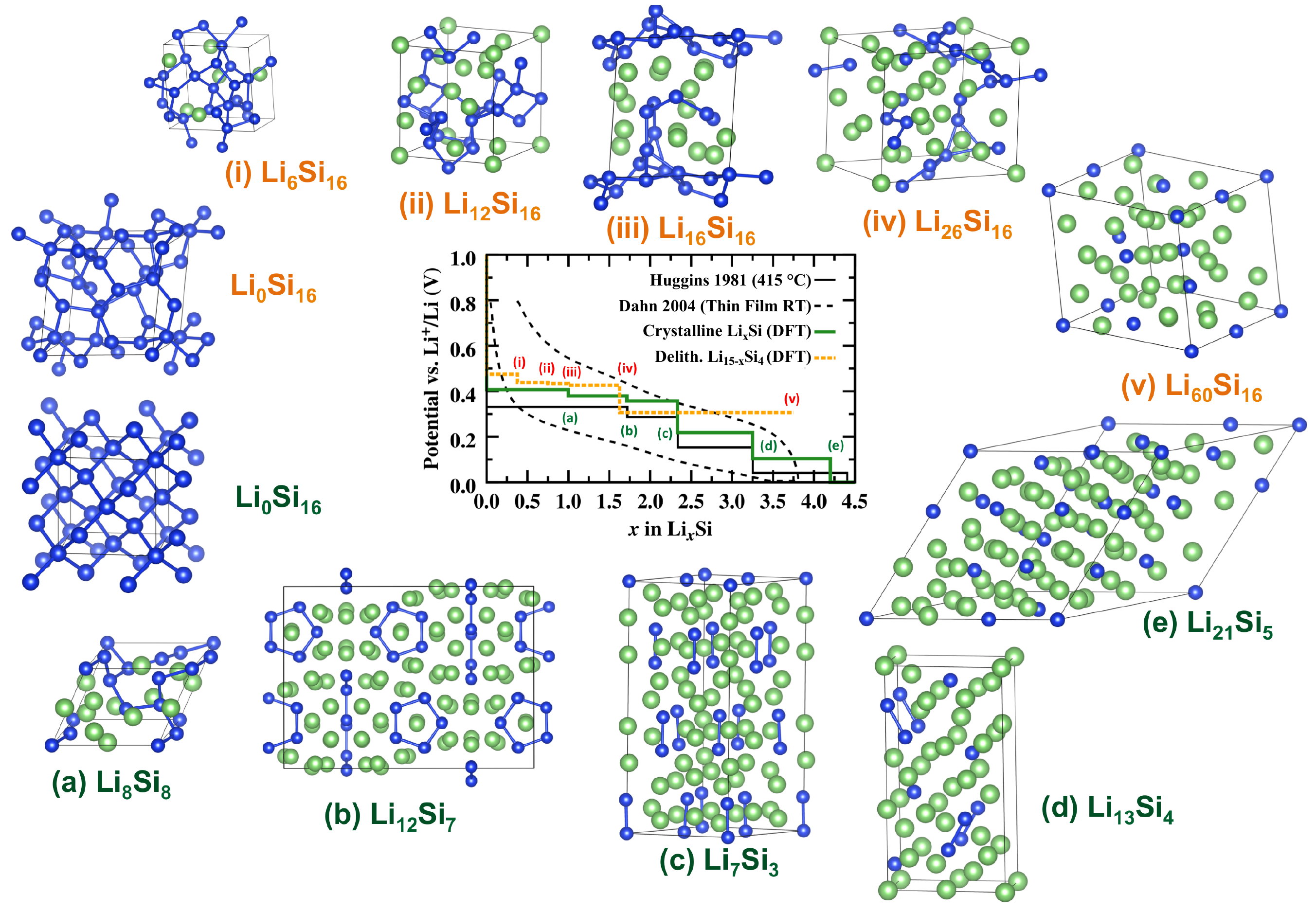}\\
\caption{
\label{fig:aLiSi-data}
\textbf{Complex phases of LiSi alloys}. Different crystalline and quasi-amorphous \ce{Li_xSi} structures (color codes: Li$=$green and Si$=$blue) that can occur during the lithiation and delithiation of silicon anodes and exhibit a wide variety of structural motifs and chemical bonds. The relative lithium content \emph{x} in the amorphous \ce{Li_xSi} alloys varies during battery charge and discharge (experimental values from Refs.~\citenum{wen_Chemical_1981} and \citenum{hatchard_Situ_2004}). The structure models were taken from Refs.~\citenum{artrith_constructing_2018} and \citenum{artrith2019a}.}
\end{figure*}

The accurate modeling of complex materials at experimental conditions often requires the use of large simulation cells containing thousands of atoms.
Long simulation times on the order of nanoseconds are needed if the dynamics of the systems are of interest.
In addition to these efficiency requirements, the complex bonding patterns that govern the atomic interactions need to be described by an appropriate method of sufficient accuracy.
MLPs trained to accurate QM calculations can fulfill both the efficiency and the accuracy requirements.
Two examples that demonstrate the need for efficient and accurate simulation tools are the transport of Li in nanostructured amorphous Si for high-energy-density batteries and diffusion phenomena in liquid water.

\textbf{Diffusion in complex amorphous \texorpdfstring{Li$_x$Si}{Li(x)Si} materials}:
Nanostructured silicon (Si) is a promising alternative to graphite as high-capacity anode material for Li-ion batteries~\cite{am25-2013-4966, aem4-2013-1300882, jmca1-2013-9566, domi_analysis_2020, domi_lithiation_2021, cangaz_enabling_2020, cao_solid_2019, zhu_minimized_2019}.
Nanoscaled materials are required to overcome mechanical limitations and avoid fracturing~\cite{am27-2015-1526}.
In addition, the rate-capability of silicon anodes is 2--4~orders of magnitude lower~\cite{ssi180-2009-222, mcp120-2010-421, jpcc116-2012-1472} compared to graphite anodes~\cite{jps195-2010-7904} owing to lower Li diffusivity.
The lithiation of crystalline Si (c-\ce{Si}) results in amorphization of the silicon crystal structure~\cite{am25-2013-4966}.
The amorphous structure of silicon anodes, the length-scale dependence of its mechanical properties, and the comparatively low lithium conductivity make the investigation of lithium transport in silicon electrodes challenging for computation and experiment.
We have previously employed ANN potentials to investigate the complex phase diagram of amorphous LiSi alloys~\cite{artrith_constructing_2018} and the delithiation of entire \ce{Li_xSi} nanoparticles~\cite{artrith2019a}.
Fig.~\ref{fig:aLiSi-data} shows examples of different crystalline and quasi-amorphous \ce{Li_xSi} structures that can occur during the lithiation and delithiation of silicon anodes, which exhibit a wide variety of structural motifs and chemical bonds.

These considerations make this system a prototypical complex material that cannot easily be modeled with either first principles methods or conventional interatomic potentials but can be successfully described by ANN potentials~\cite{artrith_constructing_2018,artrith2019a}.

\begin{figure}
\centering
  \includegraphics[scale=0.55]{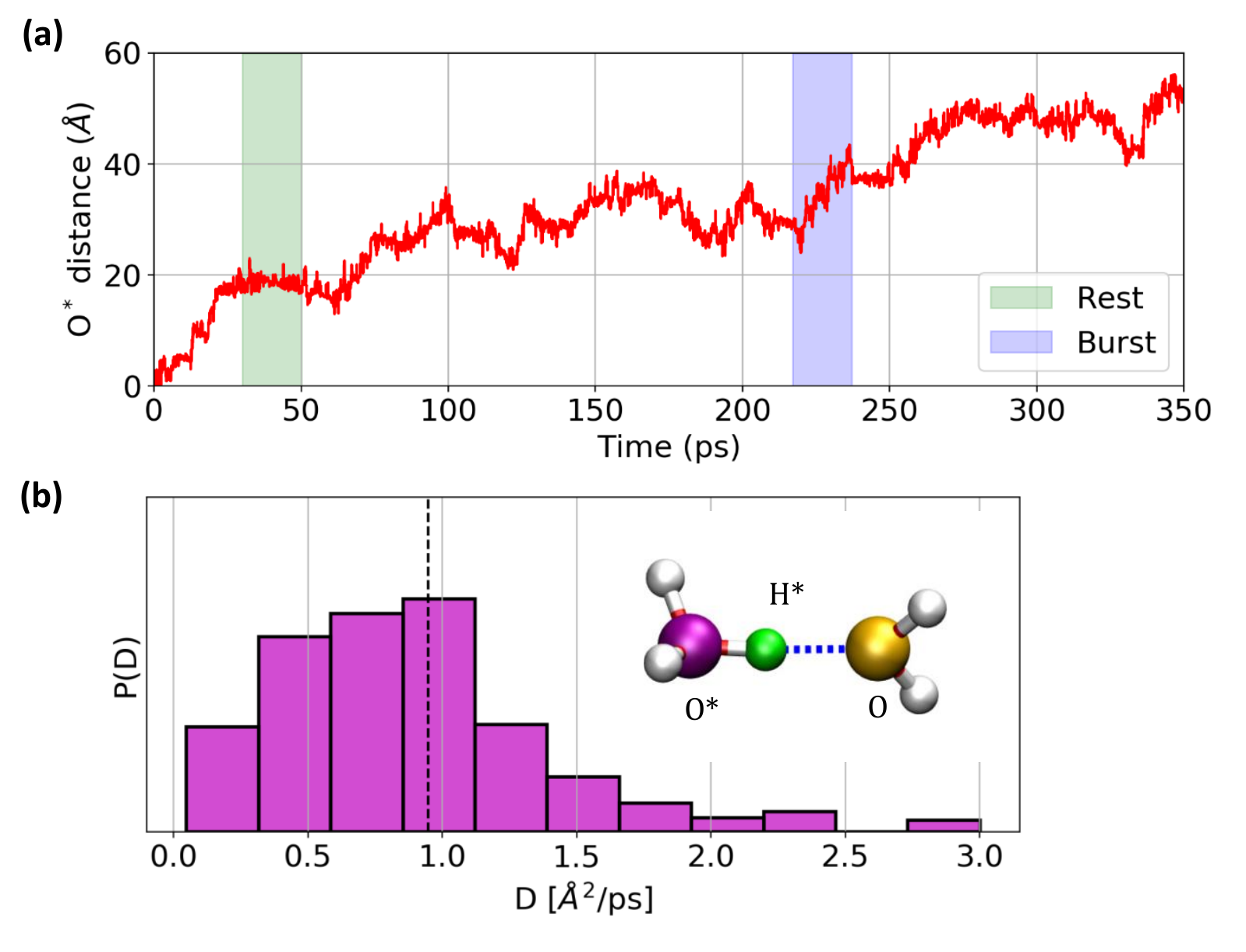}\\
\caption{
\label{fig:proton_diffusion}
\textbf{Importance of long-time simulations for predicting proton defect diffusion.}
\textbf{(a)} The diffusion process of a hydronium ($H_3O^+$) defect in liquid water, shown as the distance of the defect position at time $t_i$ from its position at time $t_0$, is characterized by alternating periods of rapid defect movement (\textit{burst} phase highlighted in blue) and little change in position ($rest$ phase highlighted in green) which can extend over 10s of picoseconds. \textbf{(b)} Defect diffusion coefficients obtained from 20 picosecond segments extracted from the 350 ps trajectory shown in (a) are distributed across a large value range indicating that a single trajectory of that length is insufficient to estimate the proton defect diffusion coefficient.
Results from periodic AIMD simulations containing 63 \ce{H2O} molecules and a single \ce{H3O} defect. \cite{napoli_decoding_2018}
}
\end{figure}

\textbf{Diffusion in liquid water}:
The need for efficient simulation tools coupled to MLPs becomes also apparent when studying diffusion phenomena in complex molecular systems.
One example is the transport of proton defects\cite{voth_Computer_2006,marx_Aqueous_2010}, a crucial process for the understanding of electrochemical reactions and enzymatic mechanisms that requires a simulation method with the ability to break and form chemical bonds.
Ab initio molecular dynamics (AIMD) simulations are in principle able to describe the diffusion of protons in hydrogen-bond networks but are limited to short time-scales and often neglect nuclear quantum effects\cite{markland2018}.
As demonstrated in Fig.~\ref{fig:proton_diffusion} short simulations with a length of 10s of picoseconds are not sufficient to reliably obtain proton defect diffusion coefficients in liquid water.
This is a consequence of the Grotthuss mechanism in which structural diffusion occurs in a stepwise fashion with periods of rapid defect diffusion followed by rest states with little movement~\cite{hassanali_Proton_2013,tse_Analysis_2015} that makes it necessary to follow the defect trajectories over several nanoseconds.
The issue is further complicated by the fact that periodic simulation cells are usually small in size to save computing resources and therefore often contain only a single defect.
Even when averaging over multiple atoms is possible, as in case of molecular diffusion in pure liquid water, an accurate prediction and comparison to the experimentally observed property is not straightforward.
While structural properties of liquid water such as pair distribution functions can be routinely obtained from direct AIMD simulations employing relatively small simulation cells and moderate simulation times, dynamical properties are more challenging since they have a strong system size dependence and require longer times to converge~\cite{kuhne2009}.

As these two examples demonstrate, efficient tools to generate long trajectories for complex atomistic system are a prerequisite for obtaining reliable molecular and materials properties.
In the next section we show how the \ae{}net-TINKER and \ae{}net-LAMMPS interfaces can meet these requirements while making efficient use of modern compute hardware distributing compute load across CPUs.


We demonstrate the computational efficiency of the combination of MLPs with the MD codes TINKER (Fig.~\ref{fig:scaling-tinker}) and LAMMPS (Fig.~\ref{fig:scaling-lammps}) for amorphous LiSi and bulk water with simulation cells of increasing size containing up to millions of atoms.
ANN potentials for both systems trained to publicly available datasets~\cite{chen2020b} are provided with the implementation together with the corresponding input files and reference output to facilitate a quick set up of the simulation and guarantee reproducibility~\cite{artrith_best_2021}.
In brief, these ANNs are trained to reference DFT calculations, for LiSi with the PBE~\cite{perdew1996} density-functional on data covering bulk, surface, and cluster structures at various LiSi compositions, see details in Refs.~\citenum{artrith_constructing_2018} and~\citenum{artrith2019a}.
The liquid water ANN was trained on liquid water structures across the full liquid temperature range described by the revPBE-D3~\cite{perdew1996,zhang1998,grimme2010} density-functional. We have previously shown that an ANN potential trained to this same dataset with a different set of descriptors yields accurate results across the full liquid temperature range~\cite{morawietz2018}.
A recent study also found that local environments sampled in liquid water are sufficiently similar to the environments in different ice phases that an MLP trained on only liquid water configurations is able reproduce many properties of ice~\cite{monserrat_Liquid_2020}.
For details on the availability of potentials and data sets see section Data Availability.

\subsection{Performance of \ae{}net-TINKER and \ae{}net-LAMMPS interfaces}
\label{sec:performance}

\begin{figure*}
\centering
  \includegraphics[scale=0.42]{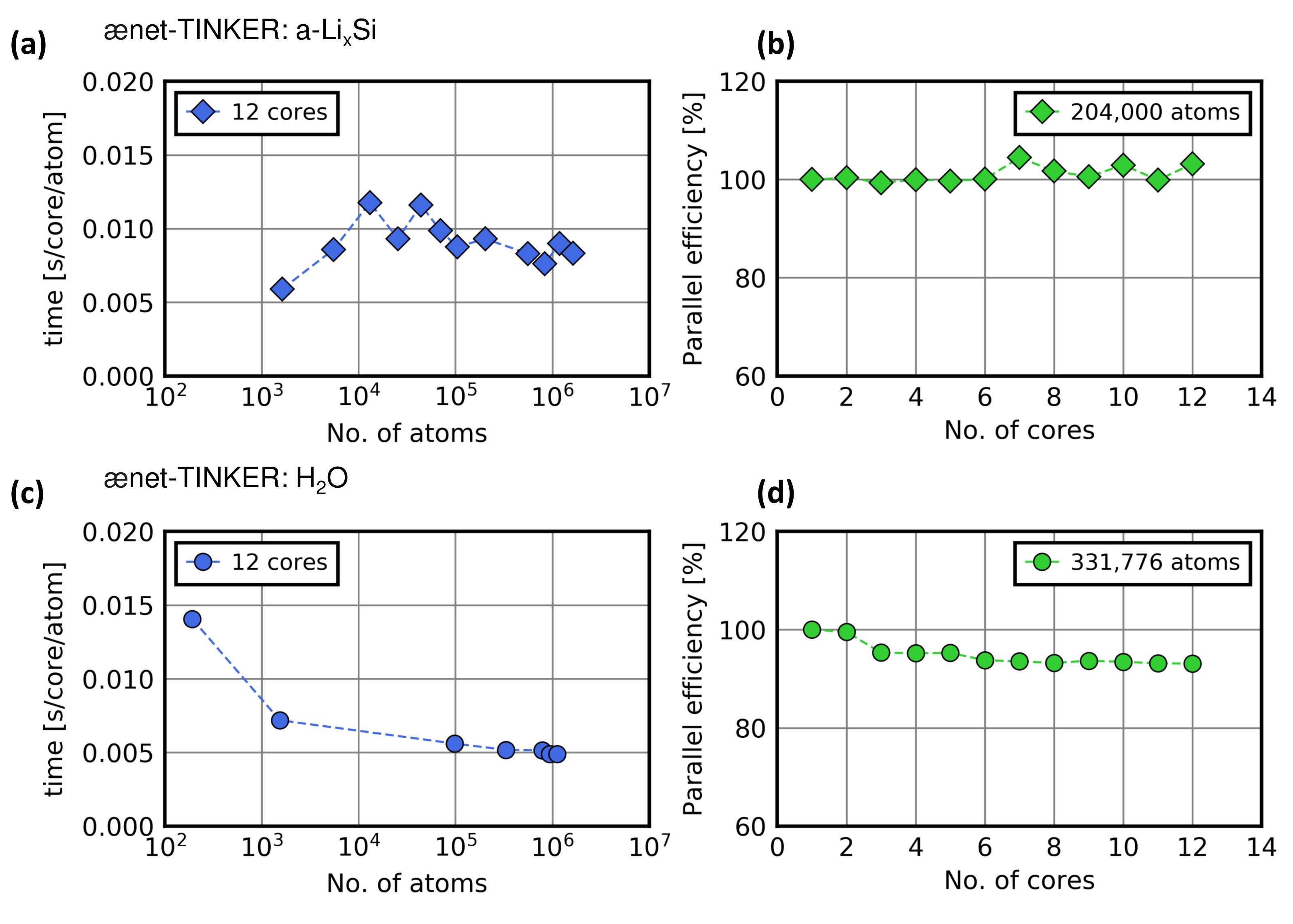}\\
\caption{
\label{fig:scaling-tinker}
\textbf{Computational efficiency of molecular dynamics simulations with the \ae{}net-TINKER interface.}
\textbf{(a-b)} \textit{a}-LiSi and \textbf{(c-d)} water. Computer cluster specifications: Dell C6420 nodes with dual Intel Xeon Gold 6126 Processor (2.6 Ghz) with Intel compiler 2017, EDR Infiniband, Red Hat Enterprise Linux 7.
}
\end{figure*}

\begin{figure*}
\centering
  \includegraphics[scale=0.42]{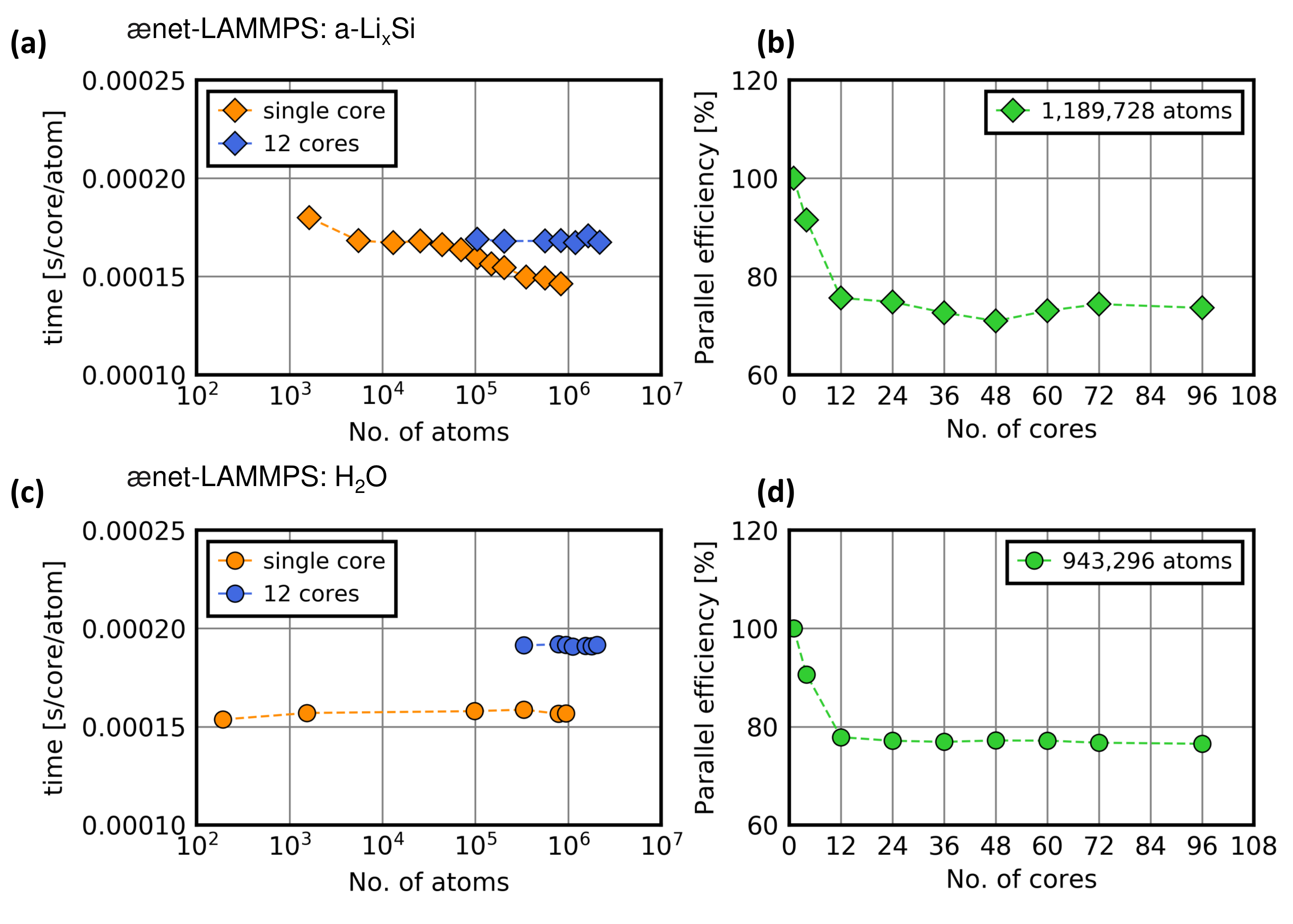}\\
\caption{
\label{fig:scaling-lammps}
\textbf{Computational efficiency of molecular dynamics simulations with the \ae{}net-LAMMPS interface.}
\textbf{(a-b)} \textit{a}-LiSi and \textbf{(c-d)} water. Computer cluster specifications: Dell C6420 nodes with dual Intel Xeon Gold 6126 Processor (2.6 Ghz) with Intel compiler 2017, EDR Infiniband, Red Hat Enterprise Linux 7.}
\end{figure*}

A major drawback of direct AIMD simulations is their steep increase in compute time with the number of atoms in the system.
Common DFT calculations have an algorithmic scaling of order $\mathcal{O}(N^3)$ or $\mathcal{O}(N\log{}N)$ with the number of electrons $N$ in the system.

To demonstrate the impact of system size on compute cost for our MLP-based simulation interfaces we compare the compute time per atom for periodic simulation cells of crystalline LiSi, panel (a) of Figs.~\ref{fig:scaling-tinker}-\ref{fig:scaling-lammps}, and bulk H$_2$O, panel (c) of Figs.~\ref{fig:scaling-tinker}-\ref{fig:scaling-lammps}, ranging from a few hundred atoms all the way up to millions of atoms.
All scaling tests were performed on the Columbia HPC cluster Terremoto, using Dell C6420 nodes (24 cores per node) with dual Intel Xeon Gold 6126 Processors (2.6 Ghz) using the Intel 2017 compiler, EDR Infiniband, and Red Hat Enterprise Linux 7.

As can be seen from the results in Figs.~\ref{fig:scaling-tinker}-\ref{fig:scaling-lammps} computational costs for both interfaces scale strictly linearly with system size, even up to very large unit cells, which is a result of the employed ANN approach in which the total atomic interactions are represented by local, atom-centered ANNs~\cite{behler2007,artrith2016}.
Even though we evaluate performance on two very different systems (an amorphous solid and a molecular liquid) both display very similar compute requirements, with single core compute times per atoms of $\sim$10 milliseconds with TINKER (Fig.~\ref{fig:scaling-tinker}) and $<$~1 millisecond for LAMMPS (Fig.~\ref{fig:scaling-lammps}).
As discussed, the current TINKER interface is based on the original implementation making use of OpenMP parallelization.
Even higher efficiency can be expected with the recently released TINKER-HP package~\cite{lagardere_TinkerHP_2018}.
To translate these numbers into compute times: to generate a 1 ns-long trajectory of a system of 1536 atoms (for H$_2$O) takes 134 h on a single compute core with \ae{}net-LAMMPS.

We note that the overall efficiency difference of the \ae{}net-TINKER and \ae{}net-LAMMPS interfaces is mainly due to the evaluation of the particle neighbor lists that are used to determined which atoms are in the local atomic environment of each atom.
The \ae{}net-TINKER interface currently uses the neighbor list implemented in \ae{}net, which is not optimized for MD simulations and is rebuilt at each step of the simulation.
In contrast, the \ae{}net-LAMMPS interface makes use of LAMMPS' highly optimized neighbor list implementation, which can reduce the computational cost for MD simulations significantly for large atomic structures.

To further reduce the wait time the compute load can be distributed across cores.
In panels (b,d) of Figs.~\ref{fig:scaling-tinker}-~\ref{fig:scaling-lammps} we report the parallel efficiency, i.e. the loss of compute power due to the distribution of tasks and the communication between the compute cores.
Due to OpenMP parallelization the \ae{}net-TINKER simulations (Fig.~\ref{fig:scaling-tinker}) have an efficiency very close to 100\% on a single node.
For \ae{}net-LAMMPS (Fig.~\ref{fig:scaling-lammps}) efficiency drops more but still stays at high values of around 70-80\% using up to 96 CPU cores on multiple nodes and with an overall better efficiency than the \ae{}net-TINKER interface.
Parallelization over multiple cores reduces the real wait time for our example system of 1536 atom to 3.6~h on 48~cores and just 1.8~h on 96~cores.

Apart from size and composition of the system of interest, the simulation performance is also dependent on the specifications of the ANN potential which in turn has a direct impact on the accuracy of the calculated properties.
In contrast to other MLP approaches, such as Gaussian Approximation Potentials~\cite{bartok2010}, the efficiency of ANN-based models does not depend on the size of the training dataset since the parameters that determine the potentials are only optimized once before the simulation.
For ANN potentials a main factor that determines both simulation efficiency and accuracy is the complexity of the model as defined by the number of nodes per layer.
The size of the input layer is given by the number of descriptors while the hidden layer size is a separate hyperparameter that has to be tuned during model construction.

\begin{figure*}
  \centering
  \includegraphics[scale=0.5]{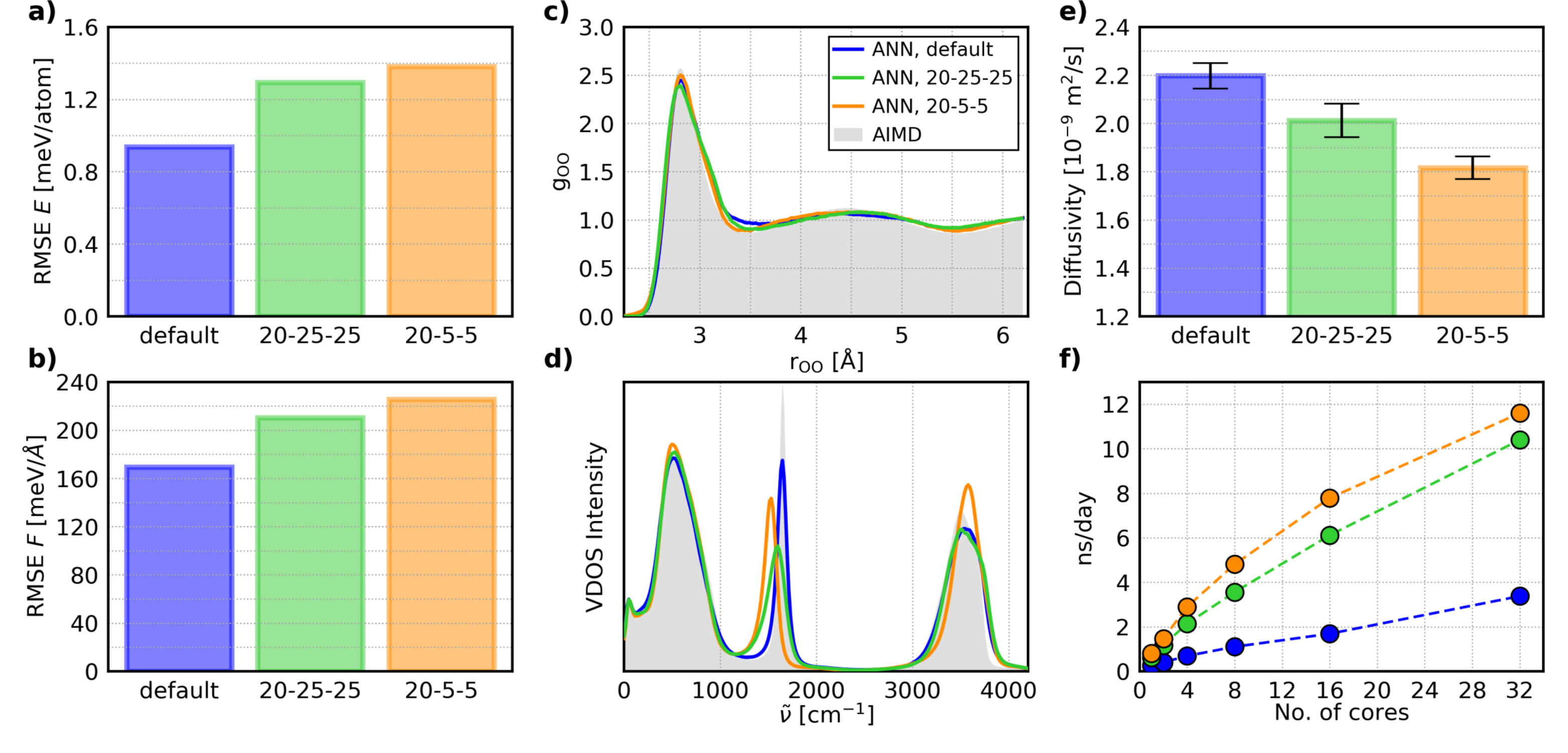}
  \caption{\label{fig:model_complexity}%
    \textbf{Impact of model complexity on accuracy and simulation efficiency}. \textbf{(a-b)}~Test set accuracy of energies (panel a) and forces (panel b) for three ANN potentials with decreasing model complexity (specified as [number of input nodes] – [number of nodes in hidden layer 1] – [number of nodes in hidden layer 2]; default model complexity: 52-25-25). \textbf{(c-d)}~Oxygen-oxygen radial distribution functions (panel c) and vibrational density of states (VDOS) of the hydrogen atoms (panel d) compared to results from direct AIMD simulations~\cite{marsalek2017}. \textbf{(e)}~System-size dependent diffusion coefficients, $D(L)$, at a box size of 64~\ce{H2O} molecules. \textbf{(f)}~Simulation time in nanoseconds accumulated on one day of compute as a function of the number of employed cores.}
\end{figure*}

To investigate these effects and provide guidance for the construction of new ANN potentials we have trained two additional potentials for water with gradually decreased complexity.
The optimal ANN potential model that was used for benchmarking the computational efficiency in Figs.~\ref{fig:scaling-tinker} and~\ref{fig:scaling-lammps} and the diffusion analysis shown in the following section has a 52-25-25-1 architecture.
First, the size of the descriptor was reduced from 52~to~20 dimensions and then, in addition, the hidden layer width was decreased from 25~to~5, i.e., the architecture was reduced to 20-25-25-1 and 20-5-5-1, respectively.
In Fig.~\ref{fig:model_complexity} the performance of the reduced models is compared to the optimal model (52-25-25-1) as a default.

As can be seen, the descriptor size is a major factor determining model accuracy (Fig.~\ref{fig:model_complexity}a--e) and simulation speed (Fig.~\ref{fig:model_complexity}f).
While cutting the descriptor leads to a simulation speed-up of a factor of~3 (Fig.~\ref{fig:model_complexity}f), the energy and force accuracy decreases (Fig.~\ref{fig:model_complexity}a--b) and deviations in dynamics properties become visible.
Although, the structural properties (as exemplified by the oxygen-oxygen RDF) show only a low variance across models (Fig.~\ref{fig:model_complexity}c), the reduced model shows a broadening and shift of the OH bending region of the hydrogen vibrational density of states (VDOS, Fig.~\ref{fig:model_complexity}d).
Further simplifying the ANN potential by reducing the hidden layer width only marginally improves the efficiency but results in greater deviations of the VDOS now also effecting the OH stretch region.

In the following, we directly demonstrate the impact of the two efficient simulation interfaces: making use of \ae{}net-TINKER to equilibrate a large nanoparticle of LiSi and using \ae{}net-LAMMPS to calculate the liquid water diffusivity at experimental conditions.

\subsection{\ae{}net-TINKER applied to LiSi}
\label{sec:application_LiSi}

\begin{figure}
\centering
  \includegraphics[scale=0.31]{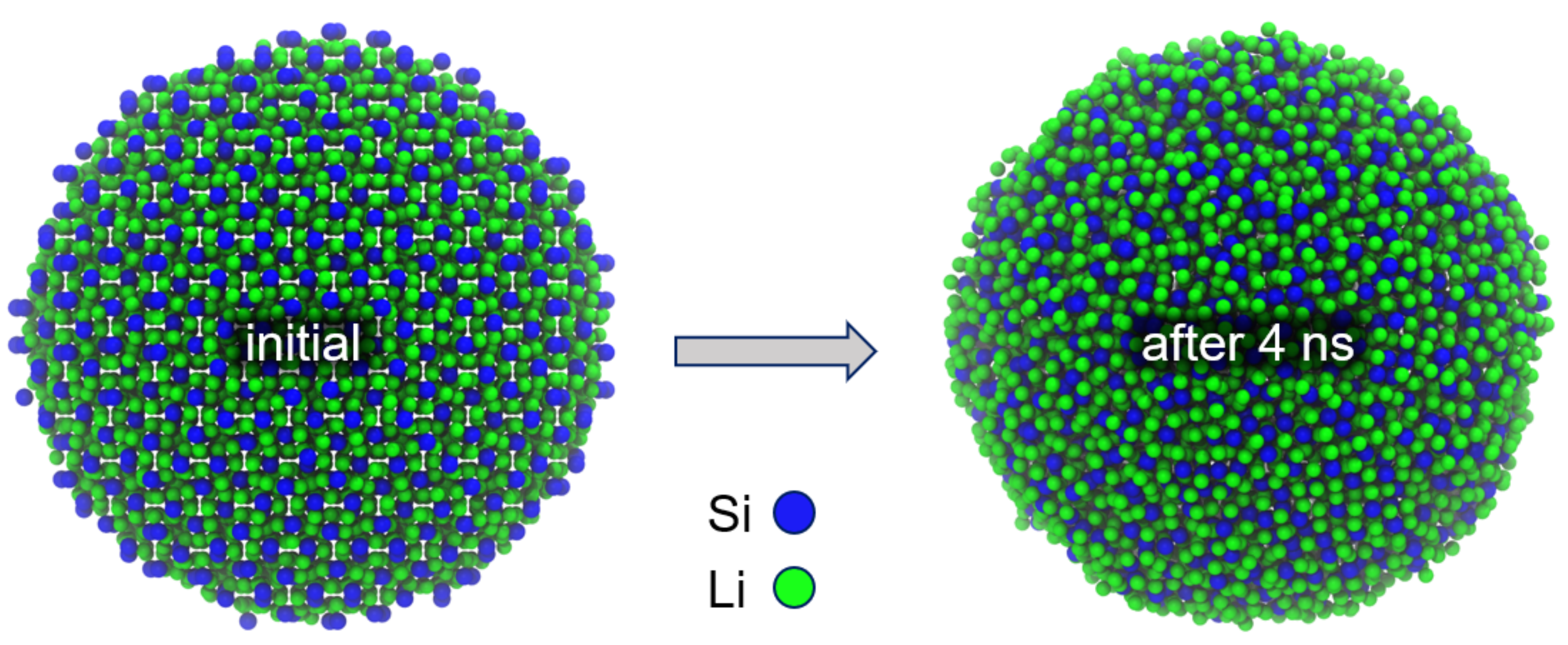}\\
\caption{
\label{fig:LiSi_equilibration_picture}
\textbf{LiSi nanoparticle equilibrated with the \ae{}net-TINKER interface.} Molecular dynamics (MD) simulation of a Li$_{8429}$Si$_{2571}$ nanoparticle structure (11,000 atoms, diameter 8 nm). Left: initial structure before MD; right: final structure after 4 ns MD simulation.}
\end{figure}

\begin{figure}
\centering
  \includegraphics[scale=0.65]{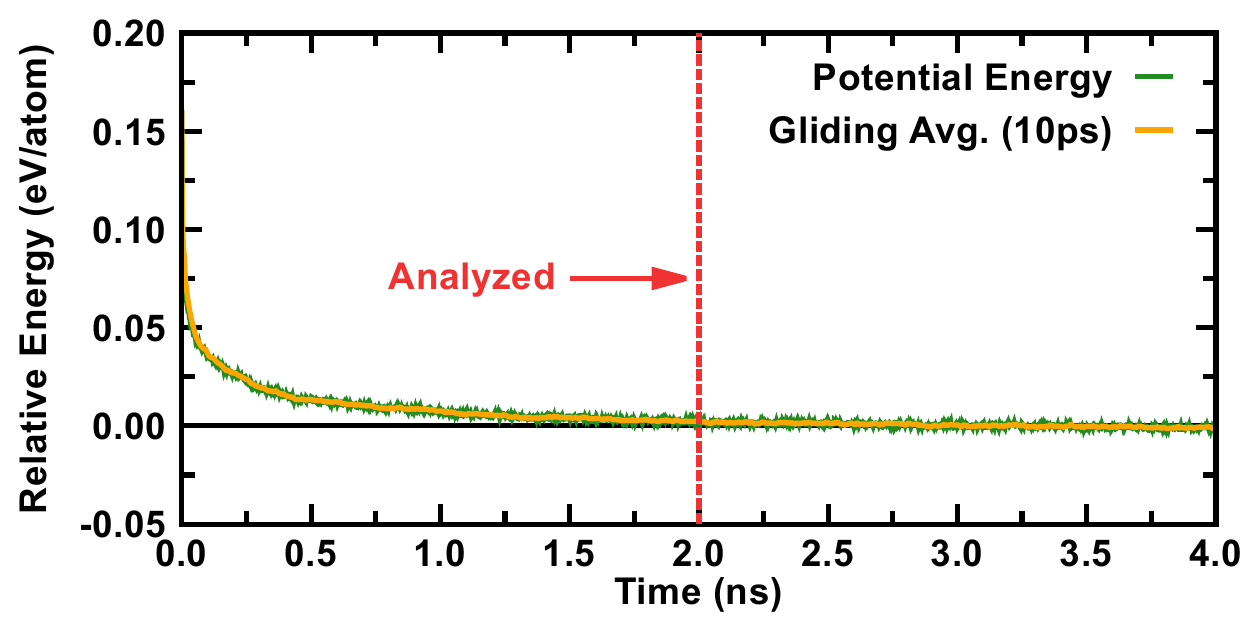}\\
\caption{
\label{fig:LiSi_equilibration_convergence}
\textbf{Equilibration of a LiSi nanoparticle with the \ae{}net-TINKER interface.} The amorphous LiSi nanoparticle structure (11,000 atoms, structure shown in Fig.~\ref{fig:LiSi_equilibration_picture}) has reached thermal equilibrium after 2 ns long  molecular dynamics (MD) simulations at 500 K.
The MD trajectory was used to analyze properties between 2 - 4 ns.
Gliding average (10 ps) of the potential energy during MD simulation of the nanoparticle model with composition Li$_{8429}$Si$_{2571}$ at 500 K.}
\end{figure}

To demonstrate the feasibility of large-scale simulations of nanostructures, we performed an MD equilibration simulation of a \ce{Li_xSi} nanoparticle with 11,000 atoms (8~nm diameter) at 500~K using the \ae{}net-TINKER interface.
The initial structure (truncated from the \ce{Li15Si4} crystal structure), and the final equilibrated structure are shown in Fig.~\ref{fig:LiSi_equilibration_picture}.
The progression of the potential energy during the MD simulation and its gliding average over 10~ps is illustrated in Fig.~\ref{fig:LiSi_equilibration_convergence}.
After around 2 ns the potential energy is within 2 meV/atom of the final structure, indicating that the system has reached thermal equilibrium.
This example further illustrates the need for efficient long-time MD simulations to investigate the properties of LiSi anodes\cite{artrith_constructing_2018, artrith2019a}.

\subsection{\ae{}net-LAMMPS applied to H$_2$O}
\label{sec:application_H2O}

As an additional demonstration of the benefits of an efficient, linear scaling MD setup for ab initio-quality simulations we report results on molecular diffusion in liquid water employing the \ae{}net-LAMMPS interface.
Diffusion coefficients were calculated from the mean square displacement of the oxygen positions
\begin{align}
  D
  = \lim_{t\rightarrow\infty} \frac{1}{6} \frac{d}{dt}
  \left< \left| r(t) - r(0) \right|^2 \right>
  \quad .
  \label{eq:diffusion_from_MSD}
\end{align}
As discussed above the accurate prediction of dynamic properties such as diffusion coefficients is particularly challenging due to their system size dependence resulting from hydrodynamic interactions~\cite{dunweg1993}.
For the small system sizes still affordable by direct AIMD this can lead do deviations between the diffusion coefficient computed at finite system size and the infinite-system limit that should be compared to the experiment~\cite{yeh2004}.
A common solution is to apply a correction formula that extrapolates the diffusion coefficient at finite system size, $D\bigl(L\bigr)$, to the infinite system size limit,
\begin{align}
  D\bigl(\infty\bigr) = D\bigl(L\bigr) + \frac{\xi}{6\pi\beta}\frac{1}{\eta L},
  \label{eq:diffusion_correction}
\end{align}
with $\beta=(k_\text{B}T)^{-1}$, the length of the cubic simulation cell $L$, the shear viscosity $\eta$, and a constant $\xi$~$=$~2.837297~\cite{dunweg1993,yeh2004}.
However this formula requires a value for the viscosity which is often taken from experiment and might not correspond to the value one would obtain from simulations with a particular computational method (density-functional, interatomic potential, or force field).

\begin{figure}
\centering
  \includegraphics[scale=0.375]{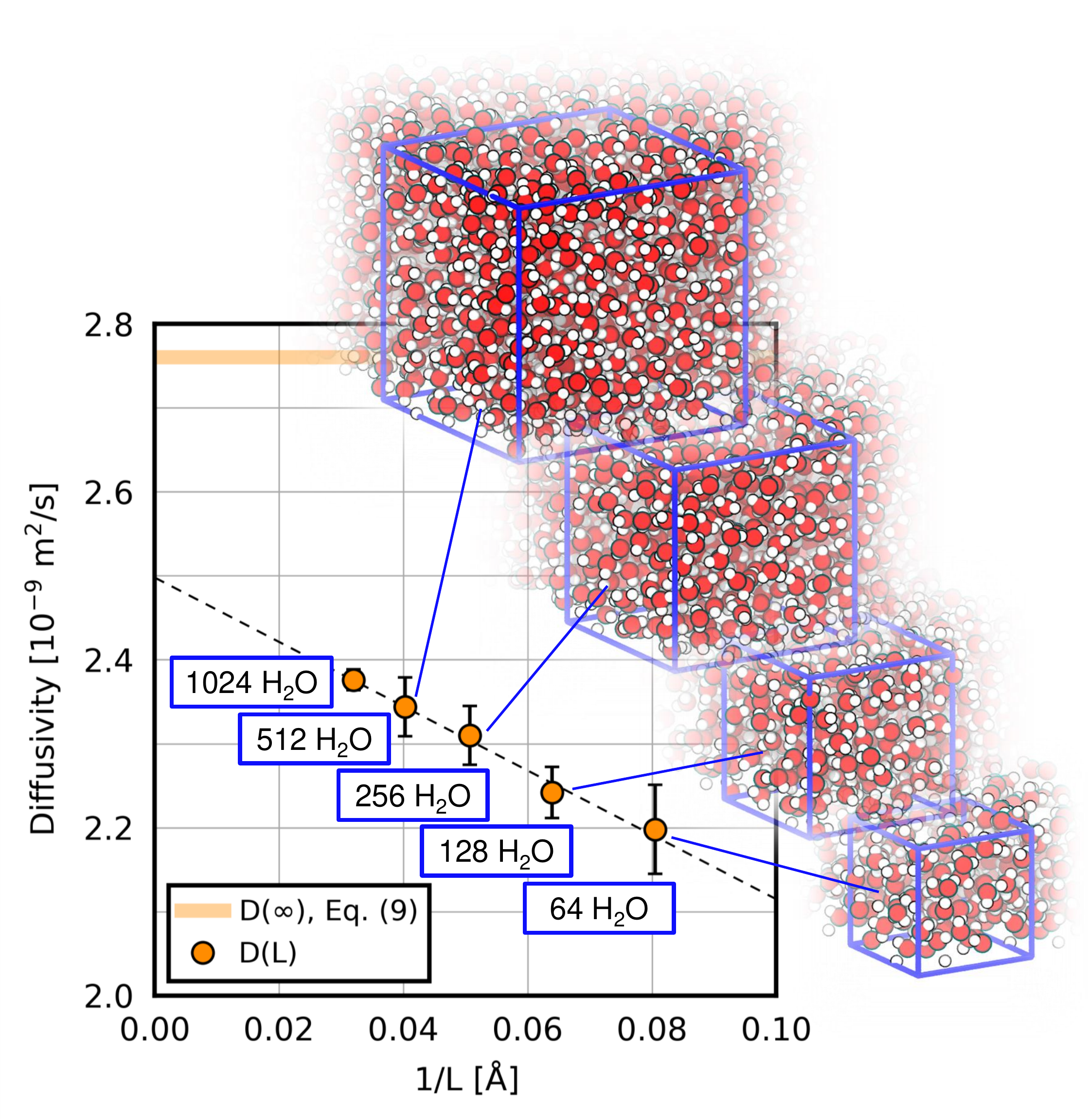}\\
\caption{
\label{fig:liquid_water_diffusion}
\textbf{Calculation of the liquid water diffusivity with the \ae{}net-LAMMPS interface.}
Liquid water diffusion coefficients as function of the inverse length of the simulation cell, $1/L$, (shown as orange circle) obtained from ANN simulations with increasingly larger simulations cells. Error bars represent the standard error of the mean from a set of diffusion coefficients obtained from five independent runs (of length 200 ps each) per system size. Extrapolation to infinite system size employing the linear relationship between diffusion and inverse box length yields an estimate of the diffusion coefficient at experimental conditions $D(\infty)=2.5\times{}10^{-9}$~m$^{2}$/s (dashed line at $1/L = 0.0$). A corrected diffusion coefficient based on the ANN results for 64~\ce{H2O} obtained with the correction formula of Eq.~\ref{eq:diffusion_correction} using the experimental viscosity value is displayed as orange shaded line.
}
\end{figure}

Using the \ae{}net-LAMMPS interface we performed MD simulations with increasingly larger simulation cells (containing up to 3072~atoms with a total time of 1 ns per system) to estimate the diffusion coefficient of liquid water at experimental conditions.
As shown in Fig.~\ref{fig:liquid_water_diffusion} the size-independent diffusion coefficient $D\bigl(\infty\bigr)$ that should be compared to experiment is obtained directly by extrapolating the diffusivity obtained from finite simulation cells to infinite system size employing the linear relationship between diffusion and inverse box length.
Compared to the estimate obtained with the correction formula, Eq.~\ref{eq:diffusion_correction}, the $D\bigl(\infty\bigr)$ value from direct extrapolation (line fit to $D(L)$ vs.\ $1/L$) is lower, indicating that the viscosity of revPBE-based ANN water is higher than the experimental one at 300~K.

\section{Conclusion}

To realize the full potential of machine learning accelerated atomistic simulations, MLPs need to be made compatible with existing highly optimized simulation software.
Here, we reported interfaces of the \ae{}net ANN potential software with the LAMMPS and TINKER simulation packages for accurate and efficient simulations of complex atomic systems.
We detailed the installation and usage and discussed implementation differences.
Both interfaces facilitate large-scale molecular dynamics simulations on modern parallel computer architectures with single step compute times of (sub-)milliseconds for both systems, an amorphous material and a molecular liquid.
The shared-memory parallelized \ae{}net-TINKER interface shows nearly perfect parallel efficiency on a single computer node.
The distributed-memory parallelized \ae{}net-LAMMPS interface scales with excellent efficiency ($\sim$80\%) on multi-node architectures as tested here with up to $\sim$100 computer cores.
The \ae{}net-LAMMPS interface further shows generally a higher efficiency than the \ae{}net-TINKER interface owing to the highly optimized neighbor list implementation in LAMMPS.
The \ae{}net-TINKER currently uses \ae{}net's internal neighbor list that is not optimized for MD simulations, and the efficiency of the interface for large structures could be further improved by making use of TINKER's own implementation.
Both interfaces make it possible to simulate atomic structures with several hundreds of thousands to millions of atoms with an accuracy close to first principles calculations.
Further improvement in the efficiency of such simulations can be expected in the future when support for graphical processor units (GPUs) becomes available.

\section*{Data availability}


\textbf{DFT reference data}: The reference datasets of the crystalline/amorphous LiSi and liquid water systems can be obtained from the Materials Cloud repository (\url{https://doi.org/10.24435/materialscloud:dx-ct}). These datasets contain atomic structures and interatomic forces in the XCrySDen~\cite{kokalj1999} structure format (XSF), and total energies are included as additional meta information.

\textbf{Code availability}: The \ae{}net source code and its documentation can be obtained either from the \ae{}net website (\url{http://ann.atomistic.net}) or from GitHub (\url{https://github.com/atomisticnet/aenet}). The implementation and examples including ANN potentials for crystalline/amorphous LiSi and liquid water can be found here:
1. \ae{}net-LAMMPS code (\url{https://github.com/atomisticnet/aenet-lammps});
2. \ae{}net-TINKER code (\url{https://github.com/atomisticnet/aenet-tinker}).

\textbf{Tutorials}: Jupyter notebooks with tutorials demonstrating the usage of the TINKER and LAMMPS interfaces for three different example systems can be found at:
\begin{itemize}
\item \ae{}net-tinker (amorphous LiSi):
\url{https://github.com/atomisticnet/aenet-tinker/tree/master/tutorial};
\item \ae{}net-lammps (liquid bulk water):
\url{https://github.com/atomisticnet/aenet-lammps/tree/master/tutorial};
\item \ae{}net-lammps (BCC iron):
\url{https://github.com/HidekiMori-CIT/aenet-lammps/tree/master/tutorial}.
\end{itemize}

\vspace{-0.5\baselineskip}

\section*{Acknowledgments}
\vspace{-0.2\baselineskip}

M.S.C., T.M., and T.E.M.\ thank Dr.~Ondrej Marsalek for discussions on the simulation of diffusion phenomena in liquid water and Austin Atsango for discussions on the proton analysis.
M.S.C and T.E.M. were supported by the U.S. Department of Energy, Office of Science, Office of Basic Energy Sciences under Award Number DE-SC0020203. T.E.M. also acknowledges support from the Camille Dreyfus Teacher-Scholar Awards Program.
T.M. is grateful for financial support by the DFG (MO 3177/1-1).
N.A.\ acknowledges the support by the Columbia Center for Computational Electrochemistry (CCCE), and is grateful for discussions with the CCCE team including Dr.~Alexander Urban, Dr.~Richard A.\ Friesner, Dr.~David R.\ Reichman from Columbia University, and Dr.~James Stevenson, Dr.~Leif Jacobson, Dr.~Steven Dajnowicz from Schr\"odinger, Inc.
N.A. thanks the Extreme Science and Engineering Discovery Environment (XSEDE), which is supported by the National Science Foundation under Grant No.\ ACI-1053575 (allocation DMR140005), for supporting the development of the Atomic Energy Network (\ae{}net) package.
AENET-LAMMPS and AENET-TINKER scaling tests used HP computing resources from Columbia University’s Shared Research Computing Facility project, which is supported by NIH Research Facility Improvement Grant 1G20RR030893-01, and associated funds from the New York State Empire State Development, Division of Science Technology and Innovation (NYSTAR) contract C090171, both awarded 15 April 2010.



\vspace{-0.5\baselineskip}
\bibliographystyle{apsrev4-1}
\bibliography{interface_paper}

\begin{thebibliography}{83}%
\makeatletter
\providecommand \@ifxundefined [1]{%
 \@ifx{#1\undefined}
}%
\providecommand \@ifnum [1]{%
 \ifnum #1\expandafter \@firstoftwo
 \else \expandafter \@secondoftwo
 \fi
}%
\providecommand \@ifx [1]{%
 \ifx #1\expandafter \@firstoftwo
 \else \expandafter \@secondoftwo
 \fi
}%
\providecommand \natexlab [1]{#1}%
\providecommand \enquote  [1]{``#1''}%
\providecommand \bibnamefont  [1]{#1}%
\providecommand \bibfnamefont [1]{#1}%
\providecommand \citenamefont [1]{#1}%
\providecommand \href@noop [0]{\@secondoftwo}%
\providecommand \href [0]{\begingroup \@sanitize@url \@href}%
\providecommand \@href[1]{\@@startlink{#1}\@@href}%
\providecommand \@@href[1]{\endgroup#1\@@endlink}%
\providecommand \@sanitize@url [0]{\catcode `\\12\catcode `\$12\catcode
  `\&12\catcode `\#12\catcode `\^12\catcode `\_12\catcode `\%12\relax}%
\providecommand \@@startlink[1]{}%
\providecommand \@@endlink[0]{}%
\providecommand \url  [0]{\begingroup\@sanitize@url \@url }%
\providecommand \@url [1]{\endgroup\@href {#1}{\urlprefix }}%
\providecommand \urlprefix  [0]{URL }%
\providecommand \Eprint [0]{\href }%
\providecommand \doibase [0]{http://dx.doi.org/}%
\providecommand \selectlanguage [0]{\@gobble}%
\providecommand \bibinfo  [0]{\@secondoftwo}%
\providecommand \bibfield  [0]{\@secondoftwo}%
\providecommand \translation [1]{[#1]}%
\providecommand \BibitemOpen [0]{}%
\providecommand \bibitemStop [0]{}%
\providecommand \bibitemNoStop [0]{.\EOS\space}%
\providecommand \EOS [0]{\spacefactor3000\relax}%
\providecommand \BibitemShut  [1]{\csname bibitem#1\endcsname}%
\let\auto@bib@innerbib\@empty
\bibitem [{\citenamefont {Frenkel}\ and\ \citenamefont
  {Smit}(2001)}]{frenkel2001}%
  \BibitemOpen
  \bibfield  {author} {\bibinfo {author} {\bibfnamefont {D.}~\bibnamefont
  {Frenkel}}\ and\ \bibinfo {author} {\bibfnamefont {B.}~\bibnamefont {Smit}},\
  }\href@noop {} {\emph {\bibinfo {title} {Understanding {{Molecular
  Simulation}}: {{From Algorithms}} to {{Applications}}}}}\ (\bibinfo
  {publisher} {{Elsevier}},\ \bibinfo {year} {2001})\BibitemShut {NoStop}%
\bibitem [{\citenamefont {Allen}\ and\ \citenamefont
  {Tildesley}(2017)}]{allen2017}%
  \BibitemOpen
  \bibfield  {author} {\bibinfo {author} {\bibfnamefont {M.~P.}\ \bibnamefont
  {Allen}}\ and\ \bibinfo {author} {\bibfnamefont {D.~J.}\ \bibnamefont
  {Tildesley}},\ }\href@noop {} {\emph {\bibinfo {title} {Computer
  {{Simulation}} of {{Liquids}}}}}\ (\bibinfo  {publisher} {{Oxford University
  Press}},\ \bibinfo {year} {2017})\BibitemShut {NoStop}%
\bibitem [{\citenamefont {Curtarolo}\ \emph {et~al.}(2013)\citenamefont
  {Curtarolo}, \citenamefont {Hart}, \citenamefont {Nardelli}, \citenamefont
  {Mingo}, \citenamefont {Sanvito},\ and\ \citenamefont
  {Levy}}]{curtarolo2013}%
  \BibitemOpen
  \bibfield  {author} {\bibinfo {author} {\bibfnamefont {S.}~\bibnamefont
  {Curtarolo}}, \bibinfo {author} {\bibfnamefont {G.~L.~W.}\ \bibnamefont
  {Hart}}, \bibinfo {author} {\bibfnamefont {M.~B.}\ \bibnamefont {Nardelli}},
  \bibinfo {author} {\bibfnamefont {N.}~\bibnamefont {Mingo}}, \bibinfo
  {author} {\bibfnamefont {S.}~\bibnamefont {Sanvito}}, \ and\ \bibinfo
  {author} {\bibfnamefont {O.}~\bibnamefont {Levy}},\ }\href {\doibase
  10.1038/nmat3568} {\bibfield  {journal} {\bibinfo  {journal} {Nature
  Materials}\ }\textbf {\bibinfo {volume} {12}},\ \bibinfo {pages} {191}
  (\bibinfo {year} {2013})}\BibitemShut {NoStop}%
\bibitem [{\citenamefont {Jain}\ \emph {et~al.}(2016)\citenamefont {Jain},
  \citenamefont {Shin},\ and\ \citenamefont {Persson}}]{jain2016}%
  \BibitemOpen
  \bibfield  {author} {\bibinfo {author} {\bibfnamefont {A.}~\bibnamefont
  {Jain}}, \bibinfo {author} {\bibfnamefont {Y.}~\bibnamefont {Shin}}, \ and\
  \bibinfo {author} {\bibfnamefont {K.~A.}\ \bibnamefont {Persson}},\ }\href
  {\doibase 10.1038/natrevmats.2015.4} {\bibfield  {journal} {\bibinfo
  {journal} {Nature Reviews Materials}\ }\textbf {\bibinfo {volume} {1}},\
  \bibinfo {pages} {1} (\bibinfo {year} {2016})}\BibitemShut {NoStop}%
\bibitem [{\citenamefont {Urban}\ \emph {et~al.}(2016)\citenamefont {Urban},
  \citenamefont {Seo},\ and\ \citenamefont {Ceder}}]{urban2016}%
  \BibitemOpen
  \bibfield  {author} {\bibinfo {author} {\bibfnamefont {A.}~\bibnamefont
  {Urban}}, \bibinfo {author} {\bibfnamefont {D.-H.}\ \bibnamefont {Seo}}, \
  and\ \bibinfo {author} {\bibfnamefont {G.}~\bibnamefont {Ceder}},\ }\href
  {\doibase 10.1038/npjcompumats.2016.2} {\bibfield  {journal} {\bibinfo
  {journal} {npj Computational Materials}\ }\textbf {\bibinfo {volume} {2}},\
  \bibinfo {pages} {1} (\bibinfo {year} {2016})}\BibitemShut {NoStop}%
\bibitem [{\citenamefont {Seh}\ \emph {et~al.}(2017)\citenamefont {Seh},
  \citenamefont {Kibsgaard}, \citenamefont {Dickens}, \citenamefont
  {Chorkendorff}, \citenamefont {N{\o}rskov},\ and\ \citenamefont
  {Jaramillo}}]{seh2017}%
  \BibitemOpen
  \bibfield  {author} {\bibinfo {author} {\bibfnamefont {Z.~W.}\ \bibnamefont
  {Seh}}, \bibinfo {author} {\bibfnamefont {J.}~\bibnamefont {Kibsgaard}},
  \bibinfo {author} {\bibfnamefont {C.~F.}\ \bibnamefont {Dickens}}, \bibinfo
  {author} {\bibfnamefont {I.}~\bibnamefont {Chorkendorff}}, \bibinfo {author}
  {\bibfnamefont {J.~K.}\ \bibnamefont {N{\o}rskov}}, \ and\ \bibinfo {author}
  {\bibfnamefont {T.~F.}\ \bibnamefont {Jaramillo}},\ }\href {\doibase
  10.1126/science.aad4998} {\bibfield  {journal} {\bibinfo  {journal}
  {Science}\ }\textbf {\bibinfo {volume} {355}} (\bibinfo {year} {2017}),\
  10.1126/science.aad4998}\BibitemShut {NoStop}%
\bibitem [{\citenamefont {Oganov}\ \emph {et~al.}(2019)\citenamefont {Oganov},
  \citenamefont {Pickard}, \citenamefont {Zhu},\ and\ \citenamefont
  {Needs}}]{oganov2019}%
  \BibitemOpen
  \bibfield  {author} {\bibinfo {author} {\bibfnamefont {A.~R.}\ \bibnamefont
  {Oganov}}, \bibinfo {author} {\bibfnamefont {C.~J.}\ \bibnamefont {Pickard}},
  \bibinfo {author} {\bibfnamefont {Q.}~\bibnamefont {Zhu}}, \ and\ \bibinfo
  {author} {\bibfnamefont {R.~J.}\ \bibnamefont {Needs}},\ }\href {\doibase
  10.1038/s41578-019-0101-8} {\bibfield  {journal} {\bibinfo  {journal} {Nature
  Reviews Materials}\ }\textbf {\bibinfo {volume} {4}},\ \bibinfo {pages} {331}
  (\bibinfo {year} {2019})}\BibitemShut {NoStop}%
\bibitem [{\citenamefont {Jorgensen}(2004)}]{jorgensen2004}%
  \BibitemOpen
  \bibfield  {author} {\bibinfo {author} {\bibfnamefont {W.~L.}\ \bibnamefont
  {Jorgensen}},\ }\href {\doibase 10.1126/science.1096361} {\bibfield
  {journal} {\bibinfo  {journal} {Science}\ }\textbf {\bibinfo {volume}
  {303}},\ \bibinfo {pages} {1813} (\bibinfo {year} {2004})}\BibitemShut
  {NoStop}%
\bibitem [{\citenamefont {Van~Drie}(2007)}]{vandrie2007}%
  \BibitemOpen
  \bibfield  {author} {\bibinfo {author} {\bibfnamefont {J.~H.}\ \bibnamefont
  {Van~Drie}},\ }\href {\doibase 10.1007/s10822-007-9142-y} {\bibfield
  {journal} {\bibinfo  {journal} {Journal of Computer-Aided Molecular Design}\
  }\textbf {\bibinfo {volume} {21}},\ \bibinfo {pages} {591} (\bibinfo {year}
  {2007})}\BibitemShut {NoStop}%
\bibitem [{\citenamefont {Aminpour}\ \emph {et~al.}(2019)\citenamefont
  {Aminpour}, \citenamefont {Montemagno},\ and\ \citenamefont
  {Tuszynski}}]{aminpour2019}%
  \BibitemOpen
  \bibfield  {author} {\bibinfo {author} {\bibfnamefont {M.}~\bibnamefont
  {Aminpour}}, \bibinfo {author} {\bibfnamefont {C.}~\bibnamefont
  {Montemagno}}, \ and\ \bibinfo {author} {\bibfnamefont {J.~A.}\ \bibnamefont
  {Tuszynski}},\ }\href {\doibase 10.3390/molecules24091693} {\bibfield
  {journal} {\bibinfo  {journal} {Molecules}\ }\textbf {\bibinfo {volume}
  {24}},\ \bibinfo {pages} {1693} (\bibinfo {year} {2019})}\BibitemShut
  {NoStop}%
\bibitem [{\citenamefont {Morawietz}\ and\ \citenamefont
  {Artrith}(2020)}]{morawietz_Machine_2020}%
  \BibitemOpen
  \bibfield  {author} {\bibinfo {author} {\bibfnamefont {T.}~\bibnamefont
  {Morawietz}}\ and\ \bibinfo {author} {\bibfnamefont {N.}~\bibnamefont
  {Artrith}},\ }\href {\doibase 10.1007/s10822-020-00346-6} {\bibfield
  {journal} {\bibinfo  {journal} {Journal of Computer-Aided Molecular Design}\
  } (\bibinfo {year} {2020}),\ 10.1007/s10822-020-00346-6}\BibitemShut
  {NoStop}%
\bibitem [{\citenamefont {Becker}\ \emph {et~al.}(2013)\citenamefont {Becker},
  \citenamefont {Tavazza}, \citenamefont {Trautt},\ and\ \citenamefont
  {{Buarque de Macedo}}}]{becker2013}%
  \BibitemOpen
  \bibfield  {author} {\bibinfo {author} {\bibfnamefont {C.~A.}\ \bibnamefont
  {Becker}}, \bibinfo {author} {\bibfnamefont {F.}~\bibnamefont {Tavazza}},
  \bibinfo {author} {\bibfnamefont {Z.~T.}\ \bibnamefont {Trautt}}, \ and\
  \bibinfo {author} {\bibfnamefont {R.~A.}\ \bibnamefont {{Buarque de
  Macedo}}},\ }\href {\doibase 10.1016/j.cossms.2013.10.001} {\bibfield
  {journal} {\bibinfo  {journal} {Current Opinion in Solid State and Materials
  Science}\ }\bibinfo {series} {Frontiers in {{Methods}} for {{Materials
  Simulations}}},\ \textbf {\bibinfo {volume} {17}},\ \bibinfo {pages} {277}
  (\bibinfo {year} {2013})}\BibitemShut {NoStop}%
\bibitem [{\citenamefont {Jorgensen}\ and\ \citenamefont
  {{Tirado-Rives}}(2005)}]{jorgensen2005}%
  \BibitemOpen
  \bibfield  {author} {\bibinfo {author} {\bibfnamefont {W.~L.}\ \bibnamefont
  {Jorgensen}}\ and\ \bibinfo {author} {\bibfnamefont {J.}~\bibnamefont
  {{Tirado-Rives}}},\ }\href {\doibase 10.1073/pnas.0408037102} {\bibfield
  {journal} {\bibinfo  {journal} {Proceedings of the National Academy of
  Sciences}\ }\textbf {\bibinfo {volume} {102}},\ \bibinfo {pages} {6665}
  (\bibinfo {year} {2005})}\BibitemShut {NoStop}%
\bibitem [{\citenamefont {Behler}\ and\ \citenamefont
  {Parrinello}(2007)}]{behler2007}%
  \BibitemOpen
  \bibfield  {author} {\bibinfo {author} {\bibfnamefont {J.}~\bibnamefont
  {Behler}}\ and\ \bibinfo {author} {\bibfnamefont {M.}~\bibnamefont
  {Parrinello}},\ }\href {\doibase 10.1103/PhysRevLett.98.146401} {\bibfield
  {journal} {\bibinfo  {journal} {Physical Review Letters}\ }\textbf {\bibinfo
  {volume} {98}},\ \bibinfo {pages} {146401} (\bibinfo {year}
  {2007})}\BibitemShut {NoStop}%
\bibitem [{\citenamefont {Bart{\'o}k}\ \emph {et~al.}(2010)\citenamefont
  {Bart{\'o}k}, \citenamefont {Payne}, \citenamefont {Kondor},\ and\
  \citenamefont {Cs{\'a}nyi}}]{bartok2010}%
  \BibitemOpen
  \bibfield  {author} {\bibinfo {author} {\bibfnamefont {A.~P.}\ \bibnamefont
  {Bart{\'o}k}}, \bibinfo {author} {\bibfnamefont {M.~C.}\ \bibnamefont
  {Payne}}, \bibinfo {author} {\bibfnamefont {R.}~\bibnamefont {Kondor}}, \
  and\ \bibinfo {author} {\bibfnamefont {G.}~\bibnamefont {Cs{\'a}nyi}},\
  }\href {\doibase 10.1103/PhysRevLett.104.136403} {\bibfield  {journal}
  {\bibinfo  {journal} {Physical Review Letters}\ }\textbf {\bibinfo {volume}
  {104}},\ \bibinfo {pages} {136403} (\bibinfo {year} {2010})}\BibitemShut
  {NoStop}%
\bibitem [{\citenamefont {Artrith}\ \emph {et~al.}(2011)\citenamefont
  {Artrith}, \citenamefont {Morawietz},\ and\ \citenamefont
  {Behler}}]{artrith2011}%
  \BibitemOpen
  \bibfield  {author} {\bibinfo {author} {\bibfnamefont {N.}~\bibnamefont
  {Artrith}}, \bibinfo {author} {\bibfnamefont {T.}~\bibnamefont {Morawietz}},
  \ and\ \bibinfo {author} {\bibfnamefont {J.}~\bibnamefont {Behler}},\ }\href
  {\doibase 10.1103/PhysRevB.83.153101} {\bibfield  {journal} {\bibinfo
  {journal} {Physical Review B}\ }\textbf {\bibinfo {volume} {83}} (\bibinfo
  {year} {2011}),\ 10.1103/PhysRevB.83.153101}\BibitemShut {NoStop}%
\bibitem [{\citenamefont {Jose}\ \emph {et~al.}(2012)\citenamefont {Jose},
  \citenamefont {Artrith},\ and\ \citenamefont
  {Behler}}]{jose_construction_2012}%
  \BibitemOpen
  \bibfield  {author} {\bibinfo {author} {\bibfnamefont {K.~V.~J.}\
  \bibnamefont {Jose}}, \bibinfo {author} {\bibfnamefont {N.}~\bibnamefont
  {Artrith}}, \ and\ \bibinfo {author} {\bibfnamefont {J.}~\bibnamefont
  {Behler}},\ }\href {\doibase 10.1063/1.4712397} {\bibfield  {journal}
  {\bibinfo  {journal} {The Journal of Chemical Physics}\ }\textbf {\bibinfo
  {volume} {136}},\ \bibinfo {pages} {194111} (\bibinfo {year}
  {2012})}\BibitemShut {NoStop}%
\bibitem [{\citenamefont {Botu}\ and\ \citenamefont
  {Ramprasad}(2015)}]{botu2015}%
  \BibitemOpen
  \bibfield  {author} {\bibinfo {author} {\bibfnamefont {V.}~\bibnamefont
  {Botu}}\ and\ \bibinfo {author} {\bibfnamefont {R.}~\bibnamefont
  {Ramprasad}},\ }\href {\doibase 10.1002/qua.24836} {\bibfield  {journal}
  {\bibinfo  {journal} {International Journal of Quantum Chemistry}\ }\textbf
  {\bibinfo {volume} {115}},\ \bibinfo {pages} {1074} (\bibinfo {year}
  {2015})}\BibitemShut {NoStop}%
\bibitem [{\citenamefont {Artrith}\ and\ \citenamefont
  {Urban}(2016)}]{artrith2016}%
  \BibitemOpen
  \bibfield  {author} {\bibinfo {author} {\bibfnamefont {N.}~\bibnamefont
  {Artrith}}\ and\ \bibinfo {author} {\bibfnamefont {A.}~\bibnamefont
  {Urban}},\ }\href {\doibase 10.1016/j.commatsci.2015.11.047} {\bibfield
  {journal} {\bibinfo  {journal} {Computational Materials Science}\ }\textbf
  {\bibinfo {volume} {114}},\ \bibinfo {pages} {135} (\bibinfo {year}
  {2016})}\BibitemShut {NoStop}%
\bibitem [{\citenamefont {Shapeev}(2016)}]{shapeev_Moment_2016}%
  \BibitemOpen
  \bibfield  {author} {\bibinfo {author} {\bibfnamefont {A.~V.}\ \bibnamefont
  {Shapeev}},\ }\href {\doibase 10.1137/15M1054183} {\bibfield  {journal}
  {\bibinfo  {journal} {Multiscale Modeling \& Simulation}\ }\textbf {\bibinfo
  {volume} {14}},\ \bibinfo {pages} {1153} (\bibinfo {year}
  {2016})}\BibitemShut {NoStop}%
\bibitem [{\citenamefont {Khorshidi}\ and\ \citenamefont
  {Peterson}(2016)}]{khorshidi2016}%
  \BibitemOpen
  \bibfield  {author} {\bibinfo {author} {\bibfnamefont {A.}~\bibnamefont
  {Khorshidi}}\ and\ \bibinfo {author} {\bibfnamefont {A.~A.}\ \bibnamefont
  {Peterson}},\ }\href {\doibase 10.1016/j.cpc.2016.05.010} {\bibfield
  {journal} {\bibinfo  {journal} {Computer Physics Communications}\ }\textbf
  {\bibinfo {volume} {207}},\ \bibinfo {pages} {310} (\bibinfo {year}
  {2016})}\BibitemShut {NoStop}%
\bibitem [{\citenamefont {S.~Smith}\ \emph {et~al.}(2017)\citenamefont
  {S.~Smith}, \citenamefont {Isayev},\ and\ \citenamefont
  {E.~Roitberg}}]{s.smith2017}%
  \BibitemOpen
  \bibfield  {author} {\bibinfo {author} {\bibfnamefont {J.}~\bibnamefont
  {S.~Smith}}, \bibinfo {author} {\bibfnamefont {O.}~\bibnamefont {Isayev}}, \
  and\ \bibinfo {author} {\bibfnamefont {A.}~\bibnamefont {E.~Roitberg}},\
  }\href {\doibase 10.1039/C6SC05720A} {\bibfield  {journal} {\bibinfo
  {journal} {Chemical Science}\ }\textbf {\bibinfo {volume} {8}},\ \bibinfo
  {pages} {3192} (\bibinfo {year} {2017})}\BibitemShut {NoStop}%
\bibitem [{\citenamefont {Unke}\ and\ \citenamefont {Meuwly}(2018)}]{unke2018}%
  \BibitemOpen
  \bibfield  {author} {\bibinfo {author} {\bibfnamefont {O.~T.}\ \bibnamefont
  {Unke}}\ and\ \bibinfo {author} {\bibfnamefont {M.}~\bibnamefont {Meuwly}},\
  }\href {\doibase 10.1063/1.5017898} {\bibfield  {journal} {\bibinfo
  {journal} {The Journal of Chemical Physics}\ }\textbf {\bibinfo {volume}
  {148}},\ \bibinfo {pages} {241708} (\bibinfo {year} {2018})}\BibitemShut
  {NoStop}%
\bibitem [{\citenamefont {Sch{\"u}tt}\ \emph {et~al.}(2018)\citenamefont
  {Sch{\"u}tt}, \citenamefont {Sauceda}, \citenamefont {Kindermans},
  \citenamefont {Tkatchenko},\ and\ \citenamefont {M{\"u}ller}}]{schutt2018}%
  \BibitemOpen
  \bibfield  {author} {\bibinfo {author} {\bibfnamefont {K.~T.}\ \bibnamefont
  {Sch{\"u}tt}}, \bibinfo {author} {\bibfnamefont {H.~E.}\ \bibnamefont
  {Sauceda}}, \bibinfo {author} {\bibfnamefont {P.-J.}\ \bibnamefont
  {Kindermans}}, \bibinfo {author} {\bibfnamefont {A.}~\bibnamefont
  {Tkatchenko}}, \ and\ \bibinfo {author} {\bibfnamefont {K.-R.}\ \bibnamefont
  {M{\"u}ller}},\ }\href {\doibase 10.1063/1.5019779} {\bibfield  {journal}
  {\bibinfo  {journal} {The Journal of Chemical Physics}\ }\textbf {\bibinfo
  {volume} {148}},\ \bibinfo {pages} {241722} (\bibinfo {year}
  {2018})}\BibitemShut {NoStop}%
\bibitem [{\citenamefont {Mori}\ and\ \citenamefont
  {Ozaki}(2020)}]{mori_Neural_2020}%
  \BibitemOpen
  \bibfield  {author} {\bibinfo {author} {\bibfnamefont {H.}~\bibnamefont
  {Mori}}\ and\ \bibinfo {author} {\bibfnamefont {T.}~\bibnamefont {Ozaki}},\
  }\href {\doibase 10.1103/PhysRevMaterials.4.040601} {\bibfield  {journal}
  {\bibinfo  {journal} {Physical Review Materials}\ }\textbf {\bibinfo {volume}
  {4}},\ \bibinfo {pages} {040601} (\bibinfo {year} {2020})}\BibitemShut
  {NoStop}%
\bibitem [{\citenamefont {Miksch}\ \emph {et~al.}(2021)\citenamefont {Miksch},
  \citenamefont {Morawietz}, \citenamefont {K{\"a}stner}, \citenamefont
  {Urban},\ and\ \citenamefont {Artrith}}]{miksch_Strategies_2021}%
  \BibitemOpen
  \bibfield  {author} {\bibinfo {author} {\bibfnamefont {A.~M.}\ \bibnamefont
  {Miksch}}, \bibinfo {author} {\bibfnamefont {T.}~\bibnamefont {Morawietz}},
  \bibinfo {author} {\bibfnamefont {J.}~\bibnamefont {K{\"a}stner}}, \bibinfo
  {author} {\bibfnamefont {A.}~\bibnamefont {Urban}}, \ and\ \bibinfo {author}
  {\bibfnamefont {N.}~\bibnamefont {Artrith}},\ }\href {\doibase
  10.1088/2632-2153/abfd96} {\bibfield  {journal} {\bibinfo  {journal} {Machine
  Learning: Science and Technology}\ } (\bibinfo {year} {2021}),\
  10.1088/2632-2153/abfd96}\BibitemShut {NoStop}%
\bibitem [{\citenamefont {Burke}(2012)}]{burke2012}%
  \BibitemOpen
  \bibfield  {author} {\bibinfo {author} {\bibfnamefont {K.}~\bibnamefont
  {Burke}},\ }\href {\doibase 10.1063/1.4704546} {\bibfield  {journal}
  {\bibinfo  {journal} {The Journal of Chemical Physics}\ }\textbf {\bibinfo
  {volume} {136}},\ \bibinfo {pages} {150901} (\bibinfo {year}
  {2012})}\BibitemShut {NoStop}%
\bibitem [{\citenamefont {Artrith}\ \emph {et~al.}(2013)\citenamefont
  {Artrith}, \citenamefont {Hiller},\ and\ \citenamefont
  {Behler}}]{artrith_neural_2013}%
  \BibitemOpen
  \bibfield  {author} {\bibinfo {author} {\bibfnamefont {N.}~\bibnamefont
  {Artrith}}, \bibinfo {author} {\bibfnamefont {B.}~\bibnamefont {Hiller}}, \
  and\ \bibinfo {author} {\bibfnamefont {J.}~\bibnamefont {Behler}},\ }\href
  {\doibase 10.1002/pssb.201248370} {\bibfield  {journal} {\bibinfo  {journal}
  {physica status solidi (b)}\ }\textbf {\bibinfo {volume} {250}},\ \bibinfo
  {pages} {1191} (\bibinfo {year} {2013})}\BibitemShut {NoStop}%
\bibitem [{\citenamefont {Artrith}(2019)}]{artrith2019}%
  \BibitemOpen
  \bibfield  {author} {\bibinfo {author} {\bibfnamefont {N.}~\bibnamefont
  {Artrith}},\ }\href {\doibase 10.1088/2515-7655/ab2060} {\bibfield  {journal}
  {\bibinfo  {journal} {Journal of Physics: Energy}\ }\textbf {\bibinfo
  {volume} {1}},\ \bibinfo {pages} {032002} (\bibinfo {year}
  {2019})}\BibitemShut {NoStop}%
\bibitem [{\citenamefont {Mueller}\ \emph {et~al.}(2020)\citenamefont
  {Mueller}, \citenamefont {Hernandez},\ and\ \citenamefont
  {Wang}}]{mueller2020}%
  \BibitemOpen
  \bibfield  {author} {\bibinfo {author} {\bibfnamefont {T.}~\bibnamefont
  {Mueller}}, \bibinfo {author} {\bibfnamefont {A.}~\bibnamefont {Hernandez}},
  \ and\ \bibinfo {author} {\bibfnamefont {C.}~\bibnamefont {Wang}},\ }\href
  {\doibase 10.1063/1.5126336} {\bibfield  {journal} {\bibinfo  {journal} {The
  Journal of Chemical Physics}\ }\textbf {\bibinfo {volume} {152}},\ \bibinfo
  {pages} {050902} (\bibinfo {year} {2020})}\BibitemShut {NoStop}%
\bibitem [{\citenamefont {Ponder}\ and\ \citenamefont
  {Richards}(1987)}]{ponder1987}%
  \BibitemOpen
  \bibfield  {author} {\bibinfo {author} {\bibfnamefont {J.~W.}\ \bibnamefont
  {Ponder}}\ and\ \bibinfo {author} {\bibfnamefont {F.~M.}\ \bibnamefont
  {Richards}},\ }\href {\doibase 10.1002/jcc.540080710} {\bibfield  {journal}
  {\bibinfo  {journal} {Journal of Computational Chemistry}\ }\textbf {\bibinfo
  {volume} {8}},\ \bibinfo {pages} {1016} (\bibinfo {year} {1987})}\BibitemShut
  {NoStop}%
\bibitem [{\citenamefont {Plimpton}(1995)}]{plimpton1995}%
  \BibitemOpen
  \bibfield  {author} {\bibinfo {author} {\bibfnamefont {S.}~\bibnamefont
  {Plimpton}},\ }\href {\doibase 10.1006/jcph.1995.1039} {\bibfield  {journal}
  {\bibinfo  {journal} {Journal of Computational Physics}\ }\textbf {\bibinfo
  {volume} {117}},\ \bibinfo {pages} {1} (\bibinfo {year} {1995})}\BibitemShut
  {NoStop}%
\bibitem [{\citenamefont {Thompson}\ \emph {et~al.}(2015)\citenamefont
  {Thompson}, \citenamefont {Swiler}, \citenamefont {Trott}, \citenamefont
  {Foiles},\ and\ \citenamefont {Tucker}}]{thompson2015}%
  \BibitemOpen
  \bibfield  {author} {\bibinfo {author} {\bibfnamefont {A.~P.}\ \bibnamefont
  {Thompson}}, \bibinfo {author} {\bibfnamefont {L.~P.}\ \bibnamefont
  {Swiler}}, \bibinfo {author} {\bibfnamefont {C.~R.}\ \bibnamefont {Trott}},
  \bibinfo {author} {\bibfnamefont {S.~M.}\ \bibnamefont {Foiles}}, \ and\
  \bibinfo {author} {\bibfnamefont {G.~J.}\ \bibnamefont {Tucker}},\ }\href
  {\doibase 10.1016/j.jcp.2014.12.018} {\bibfield  {journal} {\bibinfo
  {journal} {Journal of Computational Physics}\ }\textbf {\bibinfo {volume}
  {285}},\ \bibinfo {pages} {316} (\bibinfo {year} {2015})}\BibitemShut
  {NoStop}%
\bibitem [{\citenamefont {Zuo}\ \emph {et~al.}(2020)\citenamefont {Zuo},
  \citenamefont {Chen}, \citenamefont {Li}, \citenamefont {Deng}, \citenamefont
  {Chen}, \citenamefont {Behler}, \citenamefont {Cs{\'a}nyi}, \citenamefont
  {Shapeev}, \citenamefont {Thompson}, \citenamefont {Wood},\ and\
  \citenamefont {Ong}}]{zuo2020}%
  \BibitemOpen
  \bibfield  {author} {\bibinfo {author} {\bibfnamefont {Y.}~\bibnamefont
  {Zuo}}, \bibinfo {author} {\bibfnamefont {C.}~\bibnamefont {Chen}}, \bibinfo
  {author} {\bibfnamefont {X.}~\bibnamefont {Li}}, \bibinfo {author}
  {\bibfnamefont {Z.}~\bibnamefont {Deng}}, \bibinfo {author} {\bibfnamefont
  {Y.}~\bibnamefont {Chen}}, \bibinfo {author} {\bibfnamefont {J.}~\bibnamefont
  {Behler}}, \bibinfo {author} {\bibfnamefont {G.}~\bibnamefont {Cs{\'a}nyi}},
  \bibinfo {author} {\bibfnamefont {A.~V.}\ \bibnamefont {Shapeev}}, \bibinfo
  {author} {\bibfnamefont {A.~P.}\ \bibnamefont {Thompson}}, \bibinfo {author}
  {\bibfnamefont {M.~A.}\ \bibnamefont {Wood}}, \ and\ \bibinfo {author}
  {\bibfnamefont {S.~P.}\ \bibnamefont {Ong}},\ }\href {\doibase
  10.1021/acs.jpca.9b08723} {\bibfield  {journal} {\bibinfo  {journal} {The
  Journal of Physical Chemistry A}\ }\textbf {\bibinfo {volume} {124}},\
  \bibinfo {pages} {731} (\bibinfo {year} {2020})}\BibitemShut {NoStop}%
\bibitem [{\citenamefont {Artrith}\ \emph {et~al.}(2018)\citenamefont
  {Artrith}, \citenamefont {Urban},\ and\ \citenamefont
  {Ceder}}]{artrith_constructing_2018}%
  \BibitemOpen
  \bibfield  {author} {\bibinfo {author} {\bibfnamefont {N.}~\bibnamefont
  {Artrith}}, \bibinfo {author} {\bibfnamefont {A.}~\bibnamefont {Urban}}, \
  and\ \bibinfo {author} {\bibfnamefont {G.}~\bibnamefont {Ceder}},\ }\href
  {\doibase 10.1063/1.5017661} {\bibfield  {journal} {\bibinfo  {journal} {The
  Journal of Chemical Physics}\ }\textbf {\bibinfo {volume} {148}},\ \bibinfo
  {pages} {241711} (\bibinfo {year} {2018})}\BibitemShut {NoStop}%
\bibitem [{\citenamefont {Artrith}\ \emph {et~al.}(2019)\citenamefont
  {Artrith}, \citenamefont {Urban}, \citenamefont {Wang},\ and\ \citenamefont
  {Ceder}}]{artrith2019a}%
  \BibitemOpen
  \bibfield  {author} {\bibinfo {author} {\bibfnamefont {N.}~\bibnamefont
  {Artrith}}, \bibinfo {author} {\bibfnamefont {A.}~\bibnamefont {Urban}},
  \bibinfo {author} {\bibfnamefont {Y.}~\bibnamefont {Wang}}, \ and\ \bibinfo
  {author} {\bibfnamefont {G.}~\bibnamefont {Ceder}},\ }\href@noop {}
  {\bibfield  {journal} {\bibinfo  {journal} {arXiv:1901.09272 [cond-mat,
  physics:physics]}\ } (\bibinfo {year} {2019})},\ \Eprint
  {http://arxiv.org/abs/1901.09272} {arXiv:1901.09272 [cond-mat,
  physics:physics]} \BibitemShut {NoStop}%
\bibitem [{\citenamefont {Morawietz}\ \emph {et~al.}(2018)\citenamefont
  {Morawietz}, \citenamefont {Marsalek}, \citenamefont {Pattenaude},
  \citenamefont {Streacker}, \citenamefont {{Ben-Amotz}},\ and\ \citenamefont
  {Markland}}]{morawietz2018}%
  \BibitemOpen
  \bibfield  {author} {\bibinfo {author} {\bibfnamefont {T.}~\bibnamefont
  {Morawietz}}, \bibinfo {author} {\bibfnamefont {O.}~\bibnamefont {Marsalek}},
  \bibinfo {author} {\bibfnamefont {S.~R.}\ \bibnamefont {Pattenaude}},
  \bibinfo {author} {\bibfnamefont {L.~M.}\ \bibnamefont {Streacker}}, \bibinfo
  {author} {\bibfnamefont {D.}~\bibnamefont {{Ben-Amotz}}}, \ and\ \bibinfo
  {author} {\bibfnamefont {T.~E.}\ \bibnamefont {Markland}},\ }\href {\doibase
  10.1021/acs.jpclett.8b00133} {\bibfield  {journal} {\bibinfo  {journal} {The
  Journal of Physical Chemistry Letters}\ }\textbf {\bibinfo {volume} {9}},\
  \bibinfo {pages} {851} (\bibinfo {year} {2018})}\BibitemShut {NoStop}%
\bibitem [{\citenamefont {Morawietz}\ \emph {et~al.}(2019)\citenamefont
  {Morawietz}, \citenamefont {Urbina}, \citenamefont {Wise}, \citenamefont
  {Wu}, \citenamefont {Lu}, \citenamefont {{Ben-Amotz}},\ and\ \citenamefont
  {Markland}}]{morawietz2019}%
  \BibitemOpen
  \bibfield  {author} {\bibinfo {author} {\bibfnamefont {T.}~\bibnamefont
  {Morawietz}}, \bibinfo {author} {\bibfnamefont {A.~S.}\ \bibnamefont
  {Urbina}}, \bibinfo {author} {\bibfnamefont {P.~K.}\ \bibnamefont {Wise}},
  \bibinfo {author} {\bibfnamefont {X.}~\bibnamefont {Wu}}, \bibinfo {author}
  {\bibfnamefont {W.}~\bibnamefont {Lu}}, \bibinfo {author} {\bibfnamefont
  {D.}~\bibnamefont {{Ben-Amotz}}}, \ and\ \bibinfo {author} {\bibfnamefont
  {T.~E.}\ \bibnamefont {Markland}},\ }\href {\doibase
  10.1021/acs.jpclett.9b01781} {\bibfield  {journal} {\bibinfo  {journal} {The
  Journal of Physical Chemistry Letters}\ }\textbf {\bibinfo {volume} {10}},\
  \bibinfo {pages} {6067} (\bibinfo {year} {2019})}\BibitemShut {NoStop}%
\bibitem [{\citenamefont {Cybenko}(1989)}]{cybenko_approximation_1989}%
  \BibitemOpen
  \bibfield  {author} {\bibinfo {author} {\bibfnamefont {G.}~\bibnamefont
  {Cybenko}},\ }\href {\doibase 10.1007/BF02551274} {\bibfield  {journal}
  {\bibinfo  {journal} {Mathematics of Control, Signals, and Systems}\ }\textbf
  {\bibinfo {volume} {2}},\ \bibinfo {pages} {303} (\bibinfo {year}
  {1989})}\BibitemShut {NoStop}%
\bibitem [{\citenamefont {Artrith}\ \emph {et~al.}(2017)\citenamefont
  {Artrith}, \citenamefont {Urban},\ and\ \citenamefont {Ceder}}]{artrith2017}%
  \BibitemOpen
  \bibfield  {author} {\bibinfo {author} {\bibfnamefont {N.}~\bibnamefont
  {Artrith}}, \bibinfo {author} {\bibfnamefont {A.}~\bibnamefont {Urban}}, \
  and\ \bibinfo {author} {\bibfnamefont {G.}~\bibnamefont {Ceder}},\ }\href
  {\doibase 10.1103/PhysRevB.96.014112} {\bibfield  {journal} {\bibinfo
  {journal} {Physical Review B}\ }\textbf {\bibinfo {volume} {96}} (\bibinfo
  {year} {2017}),\ 10.1103/PhysRevB.96.014112}\BibitemShut {NoStop}%
\bibitem [{\citenamefont {Broyden}(1970)}]{broyden_bfgs_1970}%
  \BibitemOpen
  \bibfield  {author} {\bibinfo {author} {\bibfnamefont {C.~G.}\ \bibnamefont
  {Broyden}},\ }\href {\doibase 10.1093/imamat/6.1.76} {\bibfield  {journal}
  {\bibinfo  {journal} {{IMA} Journal of Applied Mathematics}\ }\textbf
  {\bibinfo {volume} {6}},\ \bibinfo {pages} {76} (\bibinfo {year}
  {1970})}\BibitemShut {NoStop}%
\bibitem [{\citenamefont {Fletcher}(1970)}]{fletcher_bfgs_1970}%
  \BibitemOpen
  \bibfield  {author} {\bibinfo {author} {\bibfnamefont {R.}~\bibnamefont
  {Fletcher}},\ }\href {\doibase 10.1093/comjnl/13.3.317} {\bibfield  {journal}
  {\bibinfo  {journal} {Comput. J.}\ }\textbf {\bibinfo {volume} {13}},\
  \bibinfo {pages} {317} (\bibinfo {year} {1970})}\BibitemShut {NoStop}%
\bibitem [{\citenamefont {Goldfarb}(1970)}]{goldfarb_bfgs_1970}%
  \BibitemOpen
  \bibfield  {author} {\bibinfo {author} {\bibfnamefont {D.}~\bibnamefont
  {Goldfarb}},\ }\href {\doibase 10.1090/s0025-5718-1970-0258249-6} {\bibfield
  {journal} {\bibinfo  {journal} {Math. Comp.}\ }\textbf {\bibinfo {volume}
  {24}},\ \bibinfo {pages} {23} (\bibinfo {year} {1970})}\BibitemShut {NoStop}%
\bibitem [{\citenamefont {Shanno}(1970)}]{shanno_bfgs_1970}%
  \BibitemOpen
  \bibfield  {author} {\bibinfo {author} {\bibfnamefont {D.~F.}\ \bibnamefont
  {Shanno}},\ }\href {\doibase 10.1090/s0025-5718-1970-0274029-x} {\bibfield
  {journal} {\bibinfo  {journal} {Math. Comp.}\ }\textbf {\bibinfo {volume}
  {24}},\ \bibinfo {pages} {647} (\bibinfo {year} {1970})}\BibitemShut
  {NoStop}%
\bibitem [{\citenamefont {Liu}\ and\ \citenamefont
  {Nocedal}(1989)}]{liu_lbfgs_1989}%
  \BibitemOpen
  \bibfield  {author} {\bibinfo {author} {\bibfnamefont {D.~C.}\ \bibnamefont
  {Liu}}\ and\ \bibinfo {author} {\bibfnamefont {J.}~\bibnamefont {Nocedal}},\
  }\href {\doibase 10.1007/bf01589116} {\bibfield  {journal} {\bibinfo
  {journal} {Math. Program.}\ }\textbf {\bibinfo {volume} {45}},\ \bibinfo
  {pages} {503} (\bibinfo {year} {1989})}\BibitemShut {NoStop}%
\bibitem [{\citenamefont {LeCun}\ \emph {et~al.}(2012)\citenamefont {LeCun},
  \citenamefont {Bottou}, \citenamefont {Orr},\ and\ \citenamefont
  {M{\"u}ller}}]{lecun_efficient_2012}%
  \BibitemOpen
  \bibfield  {author} {\bibinfo {author} {\bibfnamefont {Y.~A.}\ \bibnamefont
  {LeCun}}, \bibinfo {author} {\bibfnamefont {L.}~\bibnamefont {Bottou}},
  \bibinfo {author} {\bibfnamefont {G.~B.}\ \bibnamefont {Orr}}, \ and\
  \bibinfo {author} {\bibfnamefont {K.-R.}\ \bibnamefont {M{\"u}ller}},\
  }\enquote {\bibinfo {title} {Efficient backprop},}\ in\ \href {\doibase
  10.1007/978-3-642-35289-8_3} {\emph {\bibinfo {booktitle} {Neural Networks:
  Tricks of the Trade: Second Edition}}},\ \bibinfo {editor} {edited by\
  \bibinfo {editor} {\bibfnamefont {G.}~\bibnamefont {Montavon}}, \bibinfo
  {editor} {\bibfnamefont {G.~B.}\ \bibnamefont {Orr}}, \ and\ \bibinfo
  {editor} {\bibfnamefont {K.-R.}\ \bibnamefont {M{\"u}ller}}}\ (\bibinfo
  {publisher} {Springer Berlin Heidelberg},\ \bibinfo {address} {Berlin,
  Heidelberg},\ \bibinfo {year} {2012})\ pp.\ \bibinfo {pages}
  {9--48}\BibitemShut {NoStop}%
\bibitem [{\citenamefont {Larsen}\ \emph {et~al.}(2017)\citenamefont {Larsen},
  \citenamefont {Mortensen}, \citenamefont {Blomqvist}, \citenamefont
  {Castelli}, \citenamefont {Christensen}, \citenamefont
  {Du{\textbackslash}lak}, \citenamefont {Friis}, \citenamefont {Groves},
  \citenamefont {Hammer}, \citenamefont {Hargus}, \citenamefont {Hermes},
  \citenamefont {Jennings}, \citenamefont {Jensen}, \citenamefont {Kermode},
  \citenamefont {Kitchin}, \citenamefont {Kolsbjerg}, \citenamefont {Kubal},
  \citenamefont {Kaasbjerg}, \citenamefont {Lysgaard}, \citenamefont
  {Maronsson}, \citenamefont {Maxson}, \citenamefont {Olsen}, \citenamefont
  {Pastewka}, \citenamefont {Peterson}, \citenamefont {Rostgaard},
  \citenamefont {Schi{\o}tz}, \citenamefont {Sch{\"u}tt}, \citenamefont
  {Strange}, \citenamefont {Thygesen}, \citenamefont {Vegge}, \citenamefont
  {Vilhelmsen}, \citenamefont {Walter}, \citenamefont {Zeng},\ and\
  \citenamefont {Jacobsen}}]{larsen_Atomic_2017}%
  \BibitemOpen
  \bibfield  {author} {\bibinfo {author} {\bibfnamefont {A.~H.}\ \bibnamefont
  {Larsen}}, \bibinfo {author} {\bibfnamefont {J.~J.}\ \bibnamefont
  {Mortensen}}, \bibinfo {author} {\bibfnamefont {J.}~\bibnamefont
  {Blomqvist}}, \bibinfo {author} {\bibfnamefont {I.~E.}\ \bibnamefont
  {Castelli}}, \bibinfo {author} {\bibfnamefont {R.}~\bibnamefont
  {Christensen}}, \bibinfo {author} {\bibfnamefont {M.}~\bibnamefont
  {Du{\textbackslash}lak}}, \bibinfo {author} {\bibfnamefont {J.}~\bibnamefont
  {Friis}}, \bibinfo {author} {\bibfnamefont {M.~N.}\ \bibnamefont {Groves}},
  \bibinfo {author} {\bibfnamefont {B.}~\bibnamefont {Hammer}}, \bibinfo
  {author} {\bibfnamefont {C.}~\bibnamefont {Hargus}}, \bibinfo {author}
  {\bibfnamefont {E.~D.}\ \bibnamefont {Hermes}}, \bibinfo {author}
  {\bibfnamefont {P.~C.}\ \bibnamefont {Jennings}}, \bibinfo {author}
  {\bibfnamefont {P.~B.}\ \bibnamefont {Jensen}}, \bibinfo {author}
  {\bibfnamefont {J.}~\bibnamefont {Kermode}}, \bibinfo {author} {\bibfnamefont
  {J.~R.}\ \bibnamefont {Kitchin}}, \bibinfo {author} {\bibfnamefont {E.~L.}\
  \bibnamefont {Kolsbjerg}}, \bibinfo {author} {\bibfnamefont {J.}~\bibnamefont
  {Kubal}}, \bibinfo {author} {\bibfnamefont {K.}~\bibnamefont {Kaasbjerg}},
  \bibinfo {author} {\bibfnamefont {S.}~\bibnamefont {Lysgaard}}, \bibinfo
  {author} {\bibfnamefont {J.~B.}\ \bibnamefont {Maronsson}}, \bibinfo {author}
  {\bibfnamefont {T.}~\bibnamefont {Maxson}}, \bibinfo {author} {\bibfnamefont
  {T.}~\bibnamefont {Olsen}}, \bibinfo {author} {\bibfnamefont
  {L.}~\bibnamefont {Pastewka}}, \bibinfo {author} {\bibfnamefont
  {A.}~\bibnamefont {Peterson}}, \bibinfo {author} {\bibfnamefont
  {C.}~\bibnamefont {Rostgaard}}, \bibinfo {author} {\bibfnamefont
  {J.}~\bibnamefont {Schi{\o}tz}}, \bibinfo {author} {\bibfnamefont
  {O.}~\bibnamefont {Sch{\"u}tt}}, \bibinfo {author} {\bibfnamefont
  {M.}~\bibnamefont {Strange}}, \bibinfo {author} {\bibfnamefont {K.~S.}\
  \bibnamefont {Thygesen}}, \bibinfo {author} {\bibfnamefont {T.}~\bibnamefont
  {Vegge}}, \bibinfo {author} {\bibfnamefont {L.}~\bibnamefont {Vilhelmsen}},
  \bibinfo {author} {\bibfnamefont {M.}~\bibnamefont {Walter}}, \bibinfo
  {author} {\bibfnamefont {Z.}~\bibnamefont {Zeng}}, \ and\ \bibinfo {author}
  {\bibfnamefont {K.~W.}\ \bibnamefont {Jacobsen}},\ }\href {\doibase
  10.1088/1361-648X/aa680e} {\bibfield  {journal} {\bibinfo  {journal} {Journal
  of Physics: Condensed Matter}\ }\textbf {\bibinfo {volume} {29}},\ \bibinfo
  {pages} {273002} (\bibinfo {year} {2017})}\BibitemShut {NoStop}%
\bibitem [{\citenamefont {Thompson}\ \emph {et~al.}(2009)\citenamefont
  {Thompson}, \citenamefont {Plimpton},\ and\ \citenamefont
  {Mattson}}]{thompson2009general}%
  \BibitemOpen
  \bibfield  {author} {\bibinfo {author} {\bibfnamefont {A.~P.}\ \bibnamefont
  {Thompson}}, \bibinfo {author} {\bibfnamefont {S.~J.}\ \bibnamefont
  {Plimpton}}, \ and\ \bibinfo {author} {\bibfnamefont {W.}~\bibnamefont
  {Mattson}},\ }\href@noop {} {\bibfield  {journal} {\bibinfo  {journal} {The
  Journal of chemical physics}\ }\textbf {\bibinfo {volume} {131}},\ \bibinfo
  {pages} {154107} (\bibinfo {year} {2009})}\BibitemShut {NoStop}%
\bibitem [{\citenamefont {Dagum}\ and\ \citenamefont
  {Menon}(1998)}]{dagum_OpenMP_1998}%
  \BibitemOpen
  \bibfield  {author} {\bibinfo {author} {\bibfnamefont {L.}~\bibnamefont
  {Dagum}}\ and\ \bibinfo {author} {\bibfnamefont {R.}~\bibnamefont {Menon}},\
  }\href {\doibase 10.1109/99.660313} {\bibfield  {journal} {\bibinfo
  {journal} {IEEE Computational Science and Engineering}\ }\textbf {\bibinfo
  {volume} {5}},\ \bibinfo {pages} {46} (\bibinfo {year} {1998})}\BibitemShut
  {NoStop}%
\bibitem [{\citenamefont {Lagard{\`e}re}\ \emph {et~al.}(2018)\citenamefont
  {Lagard{\`e}re}, \citenamefont {Jolly}, \citenamefont {Lipparini},
  \citenamefont {Aviat}, \citenamefont {Stamm}, \citenamefont {F.~Jing},
  \citenamefont {Harger}, \citenamefont {Torabifard}, \citenamefont
  {Andr{\'e}s~Cisneros}, \citenamefont {J.~Schnieders}, \citenamefont {Gresh},
  \citenamefont {Maday}, \citenamefont {Y.~Ren}, \citenamefont {W.~Ponder},\
  and\ \citenamefont {Piquemal}}]{lagardere_TinkerHP_2018}%
  \BibitemOpen
  \bibfield  {author} {\bibinfo {author} {\bibfnamefont {L.}~\bibnamefont
  {Lagard{\`e}re}}, \bibinfo {author} {\bibfnamefont {L.-H.}\ \bibnamefont
  {Jolly}}, \bibinfo {author} {\bibfnamefont {F.}~\bibnamefont {Lipparini}},
  \bibinfo {author} {\bibfnamefont {F.}~\bibnamefont {Aviat}}, \bibinfo
  {author} {\bibfnamefont {B.}~\bibnamefont {Stamm}}, \bibinfo {author}
  {\bibfnamefont {Z.}~\bibnamefont {F.~Jing}}, \bibinfo {author} {\bibfnamefont
  {M.}~\bibnamefont {Harger}}, \bibinfo {author} {\bibfnamefont
  {H.}~\bibnamefont {Torabifard}}, \bibinfo {author} {\bibfnamefont
  {G.}~\bibnamefont {Andr{\'e}s~Cisneros}}, \bibinfo {author} {\bibfnamefont
  {M.}~\bibnamefont {J.~Schnieders}}, \bibinfo {author} {\bibfnamefont
  {N.}~\bibnamefont {Gresh}}, \bibinfo {author} {\bibfnamefont
  {Y.}~\bibnamefont {Maday}}, \bibinfo {author} {\bibfnamefont
  {P.}~\bibnamefont {Y.~Ren}}, \bibinfo {author} {\bibfnamefont
  {J.}~\bibnamefont {W.~Ponder}}, \ and\ \bibinfo {author} {\bibfnamefont
  {J.-P.}\ \bibnamefont {Piquemal}},\ }\href {\doibase 10.1039/C7SC04531J}
  {\bibfield  {journal} {\bibinfo  {journal} {Chemical Science}\ }\textbf
  {\bibinfo {volume} {9}},\ \bibinfo {pages} {956} (\bibinfo {year}
  {2018})}\BibitemShut {NoStop}%
\bibitem [{\citenamefont {Adjoua}\ \emph {et~al.}(2021)\citenamefont {Adjoua},
  \citenamefont {Lagard{\`e}re}, \citenamefont {Jolly}, \citenamefont
  {Durocher}, \citenamefont {Very}, \citenamefont {Dupays}, \citenamefont
  {Wang}, \citenamefont {Inizan}, \citenamefont {C{\'e}lerse}, \citenamefont
  {Ren}, \citenamefont {Ponder},\ and\ \citenamefont
  {Piquemal}}]{adjoua_TinkerHP_2021}%
  \BibitemOpen
  \bibfield  {author} {\bibinfo {author} {\bibfnamefont {O.}~\bibnamefont
  {Adjoua}}, \bibinfo {author} {\bibfnamefont {L.}~\bibnamefont
  {Lagard{\`e}re}}, \bibinfo {author} {\bibfnamefont {L.-H.}\ \bibnamefont
  {Jolly}}, \bibinfo {author} {\bibfnamefont {A.}~\bibnamefont {Durocher}},
  \bibinfo {author} {\bibfnamefont {T.}~\bibnamefont {Very}}, \bibinfo {author}
  {\bibfnamefont {I.}~\bibnamefont {Dupays}}, \bibinfo {author} {\bibfnamefont
  {Z.}~\bibnamefont {Wang}}, \bibinfo {author} {\bibfnamefont {T.~J.}\
  \bibnamefont {Inizan}}, \bibinfo {author} {\bibfnamefont {F.}~\bibnamefont
  {C{\'e}lerse}}, \bibinfo {author} {\bibfnamefont {P.}~\bibnamefont {Ren}},
  \bibinfo {author} {\bibfnamefont {J.~W.}\ \bibnamefont {Ponder}}, \ and\
  \bibinfo {author} {\bibfnamefont {J.-P.}\ \bibnamefont {Piquemal}},\
  }\href@noop {} {\bibfield  {journal} {\bibinfo  {journal} {arXiv:2011.01207
  [physics]}\ } (\bibinfo {year} {2021})},\ \Eprint
  {http://arxiv.org/abs/2011.01207} {arXiv:2011.01207 [physics]} \BibitemShut
  {NoStop}%
\bibitem [{\citenamefont {Wen}\ and\ \citenamefont
  {Huggins}(1981)}]{wen_Chemical_1981}%
  \BibitemOpen
  \bibfield  {author} {\bibinfo {author} {\bibfnamefont {C.~J.}\ \bibnamefont
  {Wen}}\ and\ \bibinfo {author} {\bibfnamefont {R.~A.}\ \bibnamefont
  {Huggins}},\ }\href {\doibase 10.1016/0022-4596(81)90487-4} {\bibfield
  {journal} {\bibinfo  {journal} {Journal of Solid State Chemistry}\ }\textbf
  {\bibinfo {volume} {37}},\ \bibinfo {pages} {271} (\bibinfo {year}
  {1981})}\BibitemShut {NoStop}%
\bibitem [{\citenamefont {Hatchard}\ and\ \citenamefont
  {Dahn}(2004)}]{hatchard_Situ_2004}%
  \BibitemOpen
  \bibfield  {author} {\bibinfo {author} {\bibfnamefont {T.~D.}\ \bibnamefont
  {Hatchard}}\ and\ \bibinfo {author} {\bibfnamefont {J.~R.}\ \bibnamefont
  {Dahn}},\ }\href {\doibase 10.1149/1.1739217} {\bibfield  {journal} {\bibinfo
   {journal} {Journal of The Electrochemical Society}\ }\textbf {\bibinfo
  {volume} {151}},\ \bibinfo {pages} {A838} (\bibinfo {year}
  {2004})}\BibitemShut {NoStop}%
\bibitem [{\citenamefont {McDowell}\ \emph {et~al.}(2013)\citenamefont
  {McDowell}, \citenamefont {Lee}, \citenamefont {Nix},\ and\ \citenamefont
  {Cui}}]{am25-2013-4966}%
  \BibitemOpen
  \bibfield  {author} {\bibinfo {author} {\bibfnamefont {M.~T.}\ \bibnamefont
  {McDowell}}, \bibinfo {author} {\bibfnamefont {S.~W.}\ \bibnamefont {Lee}},
  \bibinfo {author} {\bibfnamefont {W.~D.}\ \bibnamefont {Nix}}, \ and\
  \bibinfo {author} {\bibfnamefont {Y.}~\bibnamefont {Cui}},\ }\href {\doibase
  10.1002/adma.201301795} {\bibfield  {journal} {\bibinfo  {journal} {Adv.
  Mater.}\ }\textbf {\bibinfo {volume} {25}},\ \bibinfo {pages} {4966}
  (\bibinfo {year} {2013})}\BibitemShut {NoStop}%
\bibitem [{\citenamefont {Su}\ \emph {et~al.}(2013)\citenamefont {Su},
  \citenamefont {Wu}, \citenamefont {Li}, \citenamefont {Xiao}, \citenamefont
  {Lott}, \citenamefont {Lu}, \citenamefont {Sheldon},\ and\ \citenamefont
  {Wu}}]{aem4-2013-1300882}%
  \BibitemOpen
  \bibfield  {author} {\bibinfo {author} {\bibfnamefont {X.}~\bibnamefont
  {Su}}, \bibinfo {author} {\bibfnamefont {Q.}~\bibnamefont {Wu}}, \bibinfo
  {author} {\bibfnamefont {J.}~\bibnamefont {Li}}, \bibinfo {author}
  {\bibfnamefont {X.}~\bibnamefont {Xiao}}, \bibinfo {author} {\bibfnamefont
  {A.}~\bibnamefont {Lott}}, \bibinfo {author} {\bibfnamefont {W.}~\bibnamefont
  {Lu}}, \bibinfo {author} {\bibfnamefont {B.~W.}\ \bibnamefont {Sheldon}}, \
  and\ \bibinfo {author} {\bibfnamefont {J.}~\bibnamefont {Wu}},\ }\href
  {\doibase 10.1002/aenm.201300882} {\bibfield  {journal} {\bibinfo  {journal}
  {Adv. Energy Mater.}\ }\textbf {\bibinfo {volume} {4}} (\bibinfo {year}
  {2013}),\ 10.1002/aenm.201300882}\BibitemShut {NoStop}%
\bibitem [{\citenamefont {Zamfir}\ \emph {et~al.}(2013)\citenamefont {Zamfir},
  \citenamefont {Nguyen}, \citenamefont {Moyen}, \citenamefont {Lee},\ and\
  \citenamefont {Pribat}}]{jmca1-2013-9566}%
  \BibitemOpen
  \bibfield  {author} {\bibinfo {author} {\bibfnamefont {M.~R.}\ \bibnamefont
  {Zamfir}}, \bibinfo {author} {\bibfnamefont {H.~T.}\ \bibnamefont {Nguyen}},
  \bibinfo {author} {\bibfnamefont {E.}~\bibnamefont {Moyen}}, \bibinfo
  {author} {\bibfnamefont {Y.~H.}\ \bibnamefont {Lee}}, \ and\ \bibinfo
  {author} {\bibfnamefont {D.}~\bibnamefont {Pribat}},\ }\href {\doibase
  10.1039/c3ta11714f} {\bibfield  {journal} {\bibinfo  {journal} {J. Mater.
  Chem. A}\ }\textbf {\bibinfo {volume} {1}},\ \bibinfo {pages} {9566}
  (\bibinfo {year} {2013})}\BibitemShut {NoStop}%
\bibitem [{\citenamefont {Domi}\ \emph {et~al.}(2020)\citenamefont {Domi},
  \citenamefont {Usui}, \citenamefont {Ando}, \citenamefont {Nishikawa},\ and\
  \citenamefont {Sakaguchi}}]{domi_analysis_2020}%
  \BibitemOpen
  \bibfield  {author} {\bibinfo {author} {\bibfnamefont {Y.}~\bibnamefont
  {Domi}}, \bibinfo {author} {\bibfnamefont {H.}~\bibnamefont {Usui}}, \bibinfo
  {author} {\bibfnamefont {A.}~\bibnamefont {Ando}}, \bibinfo {author}
  {\bibfnamefont {K.}~\bibnamefont {Nishikawa}}, \ and\ \bibinfo {author}
  {\bibfnamefont {H.}~\bibnamefont {Sakaguchi}},\ }\href {\doibase
  10.1021/acsaem.0c01238} {\bibfield  {journal} {\bibinfo  {journal} {ACS
  Applied Energy Materials}\ }\textbf {\bibinfo {volume} {3}},\ \bibinfo
  {pages} {8619} (\bibinfo {year} {2020})}\BibitemShut {NoStop}%
\bibitem [{\citenamefont {Domi}\ \emph {et~al.}(2021)\citenamefont {Domi},
  \citenamefont {Usui}, \citenamefont {Ieuji}, \citenamefont {Nishikawa},\ and\
  \citenamefont {Sakaguchi}}]{domi_lithiation_2021}%
  \BibitemOpen
  \bibfield  {author} {\bibinfo {author} {\bibfnamefont {Y.}~\bibnamefont
  {Domi}}, \bibinfo {author} {\bibfnamefont {H.}~\bibnamefont {Usui}}, \bibinfo
  {author} {\bibfnamefont {N.}~\bibnamefont {Ieuji}}, \bibinfo {author}
  {\bibfnamefont {K.}~\bibnamefont {Nishikawa}}, \ and\ \bibinfo {author}
  {\bibfnamefont {H.}~\bibnamefont {Sakaguchi}},\ }\href {\doibase
  10.1021/acsami.0c17552} {\bibfield  {journal} {\bibinfo  {journal} {ACS
  Applied Materials \& Interfaces}\ }\textbf {\bibinfo {volume} {13}},\
  \bibinfo {pages} {3816} (\bibinfo {year} {2021})}\BibitemShut {NoStop}%
\bibitem [{\citenamefont {Cangaz}\ \emph {et~al.}(2020)\citenamefont {Cangaz},
  \citenamefont {Hippauf}, \citenamefont {Reuter}, \citenamefont {Doerfler},
  \citenamefont {Abendroth}, \citenamefont {Althues},\ and\ \citenamefont
  {Kaskel}}]{cangaz_enabling_2020}%
  \BibitemOpen
  \bibfield  {author} {\bibinfo {author} {\bibfnamefont {S.}~\bibnamefont
  {Cangaz}}, \bibinfo {author} {\bibfnamefont {F.}~\bibnamefont {Hippauf}},
  \bibinfo {author} {\bibfnamefont {F.~S.}\ \bibnamefont {Reuter}}, \bibinfo
  {author} {\bibfnamefont {S.}~\bibnamefont {Doerfler}}, \bibinfo {author}
  {\bibfnamefont {T.}~\bibnamefont {Abendroth}}, \bibinfo {author}
  {\bibfnamefont {H.}~\bibnamefont {Althues}}, \ and\ \bibinfo {author}
  {\bibfnamefont {S.}~\bibnamefont {Kaskel}},\ }\href {\doibase
  10.1002/aenm.202001320} {\bibfield  {journal} {\bibinfo  {journal} {Advanced
  Energy Materials}\ }\textbf {\bibinfo {volume} {10}},\ \bibinfo {pages}
  {2001320} (\bibinfo {year} {2020})}\BibitemShut {NoStop}%
\bibitem [{\citenamefont {Cao}\ \emph {et~al.}(2019)\citenamefont {Cao},
  \citenamefont {Abate}, \citenamefont {Sivonxay}, \citenamefont {Shyam},
  \citenamefont {Jia}, \citenamefont {Moritz}, \citenamefont {Devereaux},
  \citenamefont {Persson}, \citenamefont {Steinr{\"u}ck},\ and\ \citenamefont
  {Toney}}]{cao_solid_2019}%
  \BibitemOpen
  \bibfield  {author} {\bibinfo {author} {\bibfnamefont {C.}~\bibnamefont
  {Cao}}, \bibinfo {author} {\bibfnamefont {I.~I.}\ \bibnamefont {Abate}},
  \bibinfo {author} {\bibfnamefont {E.}~\bibnamefont {Sivonxay}}, \bibinfo
  {author} {\bibfnamefont {B.}~\bibnamefont {Shyam}}, \bibinfo {author}
  {\bibfnamefont {C.}~\bibnamefont {Jia}}, \bibinfo {author} {\bibfnamefont
  {B.}~\bibnamefont {Moritz}}, \bibinfo {author} {\bibfnamefont {T.~P.}\
  \bibnamefont {Devereaux}}, \bibinfo {author} {\bibfnamefont {K.~A.}\
  \bibnamefont {Persson}}, \bibinfo {author} {\bibfnamefont {H.-G.}\
  \bibnamefont {Steinr{\"u}ck}}, \ and\ \bibinfo {author} {\bibfnamefont
  {M.~F.}\ \bibnamefont {Toney}},\ }\href {\doibase
  10.1016/j.joule.2018.12.013} {\bibfield  {journal} {\bibinfo  {journal}
  {Joule}\ }\textbf {\bibinfo {volume} {3}},\ \bibinfo {pages} {762} (\bibinfo
  {year} {2019})}\BibitemShut {NoStop}%
\bibitem [{\citenamefont {Zhu}\ \emph {et~al.}(2019)\citenamefont {Zhu},
  \citenamefont {Liu}, \citenamefont {Lv}, \citenamefont {Mu}, \citenamefont
  {Zhao}, \citenamefont {Wang}, \citenamefont {Li}, \citenamefont {Yao},
  \citenamefont {Deng}, \citenamefont {Cui},\ and\ \citenamefont
  {Zhu}}]{zhu_minimized_2019}%
  \BibitemOpen
  \bibfield  {author} {\bibinfo {author} {\bibfnamefont {B.}~\bibnamefont
  {Zhu}}, \bibinfo {author} {\bibfnamefont {G.}~\bibnamefont {Liu}}, \bibinfo
  {author} {\bibfnamefont {G.}~\bibnamefont {Lv}}, \bibinfo {author}
  {\bibfnamefont {Y.}~\bibnamefont {Mu}}, \bibinfo {author} {\bibfnamefont
  {Y.}~\bibnamefont {Zhao}}, \bibinfo {author} {\bibfnamefont {Y.}~\bibnamefont
  {Wang}}, \bibinfo {author} {\bibfnamefont {X.}~\bibnamefont {Li}}, \bibinfo
  {author} {\bibfnamefont {P.}~\bibnamefont {Yao}}, \bibinfo {author}
  {\bibfnamefont {Y.}~\bibnamefont {Deng}}, \bibinfo {author} {\bibfnamefont
  {Y.}~\bibnamefont {Cui}}, \ and\ \bibinfo {author} {\bibfnamefont
  {J.}~\bibnamefont {Zhu}},\ }\href {\doibase 10.1126/sciadv.aax0651}
  {\bibfield  {journal} {\bibinfo  {journal} {Science Advances}\ }\textbf
  {\bibinfo {volume} {5}},\ \bibinfo {pages} {eaax0651} (\bibinfo {year}
  {2019})}\BibitemShut {NoStop}%
\bibitem [{\citenamefont {Wang}\ \emph {et~al.}(2015)\citenamefont {Wang},
  \citenamefont {Li}, \citenamefont {Luo}, \citenamefont {Hao}, \citenamefont
  {Zhou}, \citenamefont {Zhang}, \citenamefont {Fan},\ and\ \citenamefont
  {Zhi}}]{am27-2015-1526}%
  \BibitemOpen
  \bibfield  {author} {\bibinfo {author} {\bibfnamefont {B.}~\bibnamefont
  {Wang}}, \bibinfo {author} {\bibfnamefont {X.}~\bibnamefont {Li}}, \bibinfo
  {author} {\bibfnamefont {B.}~\bibnamefont {Luo}}, \bibinfo {author}
  {\bibfnamefont {L.}~\bibnamefont {Hao}}, \bibinfo {author} {\bibfnamefont
  {M.}~\bibnamefont {Zhou}}, \bibinfo {author} {\bibfnamefont {X.}~\bibnamefont
  {Zhang}}, \bibinfo {author} {\bibfnamefont {Z.}~\bibnamefont {Fan}}, \ and\
  \bibinfo {author} {\bibfnamefont {L.}~\bibnamefont {Zhi}},\ }\href {\doibase
  10.1002/adma.201405031} {\bibfield  {journal} {\bibinfo  {journal} {Adv.
  Mater.}\ }\textbf {\bibinfo {volume} {27}},\ \bibinfo {pages} {1526}
  (\bibinfo {year} {2015})}\BibitemShut {NoStop}%
\bibitem [{\citenamefont {Ding}\ \emph {et~al.}(2009)\citenamefont {Ding},
  \citenamefont {Xu}, \citenamefont {Yao}, \citenamefont {Wegner},
  \citenamefont {Fang}, \citenamefont {Chen},\ and\ \citenamefont
  {Lieberwirth}}]{ssi180-2009-222}%
  \BibitemOpen
  \bibfield  {author} {\bibinfo {author} {\bibfnamefont {N.}~\bibnamefont
  {Ding}}, \bibinfo {author} {\bibfnamefont {J.}~\bibnamefont {Xu}}, \bibinfo
  {author} {\bibfnamefont {Y.}~\bibnamefont {Yao}}, \bibinfo {author}
  {\bibfnamefont {G.}~\bibnamefont {Wegner}}, \bibinfo {author} {\bibfnamefont
  {X.}~\bibnamefont {Fang}}, \bibinfo {author} {\bibfnamefont {C.}~\bibnamefont
  {Chen}}, \ and\ \bibinfo {author} {\bibfnamefont {I.}~\bibnamefont
  {Lieberwirth}},\ }\href {\doibase 10.1016/j.ssi.2008.12.015} {\bibfield
  {journal} {\bibinfo  {journal} {Solid State Ionics}\ }\textbf {\bibinfo
  {volume} {180}},\ \bibinfo {pages} {222} (\bibinfo {year}
  {2009})}\BibitemShut {NoStop}%
\bibitem [{\citenamefont {Xie}\ \emph {et~al.}(2010)\citenamefont {Xie},
  \citenamefont {Imanishi}, \citenamefont {Zhang}, \citenamefont {Hirano},
  \citenamefont {Takeda},\ and\ \citenamefont {Yamamoto}}]{mcp120-2010-421}%
  \BibitemOpen
  \bibfield  {author} {\bibinfo {author} {\bibfnamefont {J.}~\bibnamefont
  {Xie}}, \bibinfo {author} {\bibfnamefont {N.}~\bibnamefont {Imanishi}},
  \bibinfo {author} {\bibfnamefont {T.}~\bibnamefont {Zhang}}, \bibinfo
  {author} {\bibfnamefont {A.}~\bibnamefont {Hirano}}, \bibinfo {author}
  {\bibfnamefont {Y.}~\bibnamefont {Takeda}}, \ and\ \bibinfo {author}
  {\bibfnamefont {O.}~\bibnamefont {Yamamoto}},\ }\href {\doibase
  10.1016/j.matchemphys.2009.11.031} {\bibfield  {journal} {\bibinfo  {journal}
  {Mater. Chem. Phys.}\ }\textbf {\bibinfo {volume} {120}},\ \bibinfo {pages}
  {421} (\bibinfo {year} {2010})}\BibitemShut {NoStop}%
\bibitem [{\citenamefont {Li}\ \emph {et~al.}(2012)\citenamefont {Li},
  \citenamefont {Xiao}, \citenamefont {Yang}, \citenamefont {Verbrugge},\ and\
  \citenamefont {Cheng}}]{jpcc116-2012-1472}%
  \BibitemOpen
  \bibfield  {author} {\bibinfo {author} {\bibfnamefont {J.}~\bibnamefont
  {Li}}, \bibinfo {author} {\bibfnamefont {X.}~\bibnamefont {Xiao}}, \bibinfo
  {author} {\bibfnamefont {F.}~\bibnamefont {Yang}}, \bibinfo {author}
  {\bibfnamefont {M.~W.}\ \bibnamefont {Verbrugge}}, \ and\ \bibinfo {author}
  {\bibfnamefont {Y.-T.}\ \bibnamefont {Cheng}},\ }\href {\doibase
  10.1021/jp207919q} {\bibfield  {journal} {\bibinfo  {journal} {J. Phys. Chem.
  C}\ }\textbf {\bibinfo {volume} {116}},\ \bibinfo {pages} {1472} (\bibinfo
  {year} {2012})}\BibitemShut {NoStop}%
\bibitem [{\citenamefont {Park}\ \emph {et~al.}(2010)\citenamefont {Park},
  \citenamefont {Zhang}, \citenamefont {Chung}, \citenamefont {Less},\ and\
  \citenamefont {Sastry}}]{jps195-2010-7904}%
  \BibitemOpen
  \bibfield  {author} {\bibinfo {author} {\bibfnamefont {M.}~\bibnamefont
  {Park}}, \bibinfo {author} {\bibfnamefont {X.}~\bibnamefont {Zhang}},
  \bibinfo {author} {\bibfnamefont {M.}~\bibnamefont {Chung}}, \bibinfo
  {author} {\bibfnamefont {G.~B.}\ \bibnamefont {Less}}, \ and\ \bibinfo
  {author} {\bibfnamefont {A.~M.}\ \bibnamefont {Sastry}},\ }\href {\doibase
  10.1016/j.jpowsour.2010.06.060} {\bibfield  {journal} {\bibinfo  {journal}
  {J. Power Sources}\ }\textbf {\bibinfo {volume} {195}},\ \bibinfo {pages}
  {7904} (\bibinfo {year} {2010})}\BibitemShut {NoStop}%
\bibitem [{\citenamefont {Napoli}\ \emph {et~al.}(2018)\citenamefont {Napoli},
  \citenamefont {Marsalek},\ and\ \citenamefont
  {Markland}}]{napoli_decoding_2018}%
  \BibitemOpen
  \bibfield  {author} {\bibinfo {author} {\bibfnamefont {J.~A.}\ \bibnamefont
  {Napoli}}, \bibinfo {author} {\bibfnamefont {O.}~\bibnamefont {Marsalek}}, \
  and\ \bibinfo {author} {\bibfnamefont {T.~E.}\ \bibnamefont {Markland}},\
  }\href {\doibase 10.1063/1.5023704} {\bibfield  {journal} {\bibinfo
  {journal} {The Journal of Chemical Physics}\ }\textbf {\bibinfo {volume}
  {148}},\ \bibinfo {pages} {222833} (\bibinfo {year} {2018})}\BibitemShut
  {NoStop}%
\bibitem [{\citenamefont {Voth}(2006)}]{voth_Computer_2006}%
  \BibitemOpen
  \bibfield  {author} {\bibinfo {author} {\bibfnamefont {G.~A.}\ \bibnamefont
  {Voth}},\ }\href {\doibase 10.1021/ar0402098} {\bibfield  {journal} {\bibinfo
   {journal} {Accounts of Chemical Research}\ }\textbf {\bibinfo {volume}
  {39}},\ \bibinfo {pages} {143} (\bibinfo {year} {2006})}\BibitemShut
  {NoStop}%
\bibitem [{\citenamefont {Marx}\ \emph {et~al.}(2010)\citenamefont {Marx},
  \citenamefont {Chandra},\ and\ \citenamefont
  {Tuckerman}}]{marx_Aqueous_2010}%
  \BibitemOpen
  \bibfield  {author} {\bibinfo {author} {\bibfnamefont {D.}~\bibnamefont
  {Marx}}, \bibinfo {author} {\bibfnamefont {A.}~\bibnamefont {Chandra}}, \
  and\ \bibinfo {author} {\bibfnamefont {M.~E.}\ \bibnamefont {Tuckerman}},\
  }\href {\doibase 10.1021/cr900233f} {\bibfield  {journal} {\bibinfo
  {journal} {Chemical Reviews}\ }\textbf {\bibinfo {volume} {110}},\ \bibinfo
  {pages} {2174} (\bibinfo {year} {2010})}\BibitemShut {NoStop}%
\bibitem [{\citenamefont {Markland}\ and\ \citenamefont
  {Ceriotti}(2018)}]{markland2018}%
  \BibitemOpen
  \bibfield  {author} {\bibinfo {author} {\bibfnamefont {T.~E.}\ \bibnamefont
  {Markland}}\ and\ \bibinfo {author} {\bibfnamefont {M.}~\bibnamefont
  {Ceriotti}},\ }\href {\doibase 10.1038/s41570-017-0109} {\bibfield  {journal}
  {\bibinfo  {journal} {Nature Reviews Chemistry}\ }\textbf {\bibinfo {volume}
  {2}},\ \bibinfo {pages} {1} (\bibinfo {year} {2018})}\BibitemShut {NoStop}%
\bibitem [{\citenamefont {Hassanali}\ \emph {et~al.}(2013)\citenamefont
  {Hassanali}, \citenamefont {Giberti}, \citenamefont {Cuny}, \citenamefont
  {K{\"u}hne},\ and\ \citenamefont {Parrinello}}]{hassanali_Proton_2013}%
  \BibitemOpen
  \bibfield  {author} {\bibinfo {author} {\bibfnamefont {A.}~\bibnamefont
  {Hassanali}}, \bibinfo {author} {\bibfnamefont {F.}~\bibnamefont {Giberti}},
  \bibinfo {author} {\bibfnamefont {J.}~\bibnamefont {Cuny}}, \bibinfo {author}
  {\bibfnamefont {T.~D.}\ \bibnamefont {K{\"u}hne}}, \ and\ \bibinfo {author}
  {\bibfnamefont {M.}~\bibnamefont {Parrinello}},\ }\href {\doibase
  10.1073/pnas.1306642110} {\bibfield  {journal} {\bibinfo  {journal}
  {Proceedings of the National Academy of Sciences}\ }\textbf {\bibinfo
  {volume} {110}},\ \bibinfo {pages} {13723} (\bibinfo {year}
  {2013})}\BibitemShut {NoStop}%
\bibitem [{\citenamefont {Tse}\ \emph {et~al.}(2015)\citenamefont {Tse},
  \citenamefont {Knight},\ and\ \citenamefont {Voth}}]{tse_Analysis_2015}%
  \BibitemOpen
  \bibfield  {author} {\bibinfo {author} {\bibfnamefont {Y.-L.~S.}\
  \bibnamefont {Tse}}, \bibinfo {author} {\bibfnamefont {C.}~\bibnamefont
  {Knight}}, \ and\ \bibinfo {author} {\bibfnamefont {G.~A.}\ \bibnamefont
  {Voth}},\ }\href {\doibase 10.1063/1.4905077} {\bibfield  {journal} {\bibinfo
   {journal} {The Journal of Chemical Physics}\ }\textbf {\bibinfo {volume}
  {142}},\ \bibinfo {pages} {014104} (\bibinfo {year} {2015})}\BibitemShut
  {NoStop}%
\bibitem [{\citenamefont {K{\"u}hne}\ \emph {et~al.}(2009)\citenamefont
  {K{\"u}hne}, \citenamefont {Krack},\ and\ \citenamefont
  {Parrinello}}]{kuhne2009}%
  \BibitemOpen
  \bibfield  {author} {\bibinfo {author} {\bibfnamefont {T.~D.}\ \bibnamefont
  {K{\"u}hne}}, \bibinfo {author} {\bibfnamefont {M.}~\bibnamefont {Krack}}, \
  and\ \bibinfo {author} {\bibfnamefont {M.}~\bibnamefont {Parrinello}},\
  }\href {\doibase 10.1021/ct800417q} {\bibfield  {journal} {\bibinfo
  {journal} {Journal of Chemical Theory and Computation}\ }\textbf {\bibinfo
  {volume} {5}},\ \bibinfo {pages} {235} (\bibinfo {year} {2009})}\BibitemShut
  {NoStop}%
\bibitem [{\citenamefont {Chen}\ \emph {et~al.}(2020)\citenamefont {Chen},
  \citenamefont {Morawietz}, \citenamefont {Markland},\ and\ \citenamefont
  {Artrith}}]{chen2020b}%
  \BibitemOpen
  \bibfield  {author} {\bibinfo {author} {\bibfnamefont {M.~S.}\ \bibnamefont
  {Chen}}, \bibinfo {author} {\bibfnamefont {T.}~\bibnamefont {Morawietz}},
  \bibinfo {author} {\bibfnamefont {T.~E.}\ \bibnamefont {Markland}}, \ and\
  \bibinfo {author} {\bibfnamefont {N.}~\bibnamefont {Artrith}},\ }\href
  {\doibase doi.org/10.24435/materialscloud:dx-ct} {\bibfield  {journal}
  {\bibinfo  {journal} {Materials Cloud Archive}\ }\textbf {\bibinfo {volume}
  {2020.92}} (\bibinfo {year} {2020}),\
  doi.org/10.24435/materialscloud:dx-ct}\BibitemShut {NoStop}%
\bibitem [{\citenamefont {Artrith}\ \emph {et~al.}(2021)\citenamefont
  {Artrith}, \citenamefont {Butler}, \citenamefont {Coudert}, \citenamefont
  {Han}, \citenamefont {Isayev}, \citenamefont {Jain},\ and\ \citenamefont
  {Walsh}}]{artrith_best_2021}%
  \BibitemOpen
  \bibfield  {author} {\bibinfo {author} {\bibfnamefont {N.}~\bibnamefont
  {Artrith}}, \bibinfo {author} {\bibfnamefont {K.~T.}\ \bibnamefont {Butler}},
  \bibinfo {author} {\bibfnamefont {F.-X.}\ \bibnamefont {Coudert}}, \bibinfo
  {author} {\bibfnamefont {S.}~\bibnamefont {Han}}, \bibinfo {author}
  {\bibfnamefont {O.}~\bibnamefont {Isayev}}, \bibinfo {author} {\bibfnamefont
  {A.}~\bibnamefont {Jain}}, \ and\ \bibinfo {author} {\bibfnamefont
  {A.}~\bibnamefont {Walsh}},\ }\href {\doibase 10.1038/s41557-021-00716-z}
  {\bibfield  {journal} {\bibinfo  {journal} {Nature Chemistry}\ }\textbf
  {\bibinfo {volume} {13}},\ \bibinfo {pages} {505} (\bibinfo {year}
  {2021})}\BibitemShut {NoStop}%
\bibitem [{\citenamefont {Perdew}\ \emph {et~al.}(1996)\citenamefont {Perdew},
  \citenamefont {Burke},\ and\ \citenamefont {Ernzerhof}}]{perdew1996}%
  \BibitemOpen
  \bibfield  {author} {\bibinfo {author} {\bibfnamefont {J.~P.}\ \bibnamefont
  {Perdew}}, \bibinfo {author} {\bibfnamefont {K.}~\bibnamefont {Burke}}, \
  and\ \bibinfo {author} {\bibfnamefont {M.}~\bibnamefont {Ernzerhof}},\ }\href
  {\doibase 10.1103/PhysRevLett.77.3865} {\bibfield  {journal} {\bibinfo
  {journal} {Physical Review Letters}\ }\textbf {\bibinfo {volume} {77}},\
  \bibinfo {pages} {3865} (\bibinfo {year} {1996})}\BibitemShut {NoStop}%
\bibitem [{\citenamefont {Zhang}\ and\ \citenamefont {Yang}(1998)}]{zhang1998}%
  \BibitemOpen
  \bibfield  {author} {\bibinfo {author} {\bibfnamefont {Y.}~\bibnamefont
  {Zhang}}\ and\ \bibinfo {author} {\bibfnamefont {W.}~\bibnamefont {Yang}},\
  }\href {\doibase 10.1103/PhysRevLett.80.890} {\bibfield  {journal} {\bibinfo
  {journal} {Physical Review Letters}\ }\textbf {\bibinfo {volume} {80}},\
  \bibinfo {pages} {890} (\bibinfo {year} {1998})}\BibitemShut {NoStop}%
\bibitem [{\citenamefont {Grimme}\ \emph {et~al.}(2010)\citenamefont {Grimme},
  \citenamefont {Antony}, \citenamefont {Ehrlich},\ and\ \citenamefont
  {Krieg}}]{grimme2010}%
  \BibitemOpen
  \bibfield  {author} {\bibinfo {author} {\bibfnamefont {S.}~\bibnamefont
  {Grimme}}, \bibinfo {author} {\bibfnamefont {J.}~\bibnamefont {Antony}},
  \bibinfo {author} {\bibfnamefont {S.}~\bibnamefont {Ehrlich}}, \ and\
  \bibinfo {author} {\bibfnamefont {H.}~\bibnamefont {Krieg}},\ }\href
  {\doibase 10.1063/1.3382344} {\bibfield  {journal} {\bibinfo  {journal} {The
  Journal of Chemical Physics}\ }\textbf {\bibinfo {volume} {132}},\ \bibinfo
  {pages} {154104} (\bibinfo {year} {2010})}\BibitemShut {NoStop}%
\bibitem [{\citenamefont {Monserrat}\ \emph {et~al.}(2020)\citenamefont
  {Monserrat}, \citenamefont {Brandenburg}, \citenamefont {Engel},\ and\
  \citenamefont {Cheng}}]{monserrat_Liquid_2020}%
  \BibitemOpen
  \bibfield  {author} {\bibinfo {author} {\bibfnamefont {B.}~\bibnamefont
  {Monserrat}}, \bibinfo {author} {\bibfnamefont {J.~G.}\ \bibnamefont
  {Brandenburg}}, \bibinfo {author} {\bibfnamefont {E.~A.}\ \bibnamefont
  {Engel}}, \ and\ \bibinfo {author} {\bibfnamefont {B.}~\bibnamefont
  {Cheng}},\ }\href {\doibase 10.1038/s41467-020-19606-y} {\bibfield  {journal}
  {\bibinfo  {journal} {Nature Communications}\ }\textbf {\bibinfo {volume}
  {11}},\ \bibinfo {pages} {5757} (\bibinfo {year} {2020})}\BibitemShut
  {NoStop}%
\bibitem [{\citenamefont {Marsalek}\ and\ \citenamefont
  {Markland}(2017)}]{marsalek2017}%
  \BibitemOpen
  \bibfield  {author} {\bibinfo {author} {\bibfnamefont {O.}~\bibnamefont
  {Marsalek}}\ and\ \bibinfo {author} {\bibfnamefont {T.~E.}\ \bibnamefont
  {Markland}},\ }\href {\doibase 10.1021/acs.jpclett.7b00391} {\bibfield
  {journal} {\bibinfo  {journal} {The Journal of Physical Chemistry Letters}\
  }\textbf {\bibinfo {volume} {8}},\ \bibinfo {pages} {1545} (\bibinfo {year}
  {2017})}\BibitemShut {NoStop}%
\bibitem [{\citenamefont {D{\"u}nweg}\ and\ \citenamefont
  {Kremer}(1993)}]{dunweg1993}%
  \BibitemOpen
  \bibfield  {author} {\bibinfo {author} {\bibfnamefont {B.}~\bibnamefont
  {D{\"u}nweg}}\ and\ \bibinfo {author} {\bibfnamefont {K.}~\bibnamefont
  {Kremer}},\ }\href {\doibase 10.1063/1.465445} {\bibfield  {journal}
  {\bibinfo  {journal} {The Journal of Chemical Physics}\ }\textbf {\bibinfo
  {volume} {99}},\ \bibinfo {pages} {6983} (\bibinfo {year}
  {1993})}\BibitemShut {NoStop}%
\bibitem [{\citenamefont {Yeh}\ and\ \citenamefont {Hummer}(2004)}]{yeh2004}%
  \BibitemOpen
  \bibfield  {author} {\bibinfo {author} {\bibfnamefont {I.-C.}\ \bibnamefont
  {Yeh}}\ and\ \bibinfo {author} {\bibfnamefont {G.}~\bibnamefont {Hummer}},\
  }\href {\doibase 10.1021/jp0477147} {\bibfield  {journal} {\bibinfo
  {journal} {The Journal of Physical Chemistry B}\ }\textbf {\bibinfo {volume}
  {108}},\ \bibinfo {pages} {15873} (\bibinfo {year} {2004})}\BibitemShut
  {NoStop}%
\bibitem [{\citenamefont {Kokalj}( Aug)}]{kokalj1999}%
  \BibitemOpen
  \bibfield  {author} {\bibinfo {author} {\bibfnamefont {A.}~\bibnamefont
  {Kokalj}},\ }\href {\doibase 10.1016/s1093-3263(99)00028-5} {\bibfield
  {journal} {\bibinfo  {journal} {Journal of Molecular Graphics \& Modelling}\
  }\textbf {\bibinfo {volume} {17}},\ \bibinfo {pages} {176} (\bibinfo {year}
  {1999 Jun-Aug})}\BibitemShut {NoStop}%
\end{thebibliography}%

\end{document}